\documentstyle[12pt,a4,epsfig,amstex,amssymb,caption2]{article}

\setcaptionwidth{15cm}
\renewcommand{\baselinestretch}{1.1}

\def\peppo{\(p \rightarrow e^+ \pi^0\)}
\def\mpeppo{p \rightarrow e^+ \pi^0}
\def\pnukp{\(p \rightarrow \bar{\nu} K^+\)}
\def\mpnukp{p \rightarrow \bar{\nu} K^+}

\def\Super-K{Super-Kamiokande}
\def\Hyper-K{Hyper-Kamiokande}

\begin{document}
\parindent=12pt
\title{The JHF-Kamioka neutrino project}
\author{Y.~Itow$^1$, T.~Kajita$^1$, K.~Kaneyuki$^1$, M.~Shiozawa$^1$, Y.~Totsuka$^1$, \\
Y.~Hayato$^2$, T.~Ishida$^2$, T.~Ishii$^2$, T.~Kobayashi$^2$, T.~Maruyama$^2$, \\ 
K.~Nakamura$^2$, Y.~Obayashi$^2$, Y.~Oyama$^2$, M.~Sakuda$^2$, M.~Yoshida$^2$, \\ 
S.~Aoki$^3$, T.~Hara$^3$, A.~Suzuki$^3$, \\ 
A.~Ichikawa$^4$, T.~Nakaya$^4$, K.~Nishikawa$^4$, \\
T.~Hasegawa$^5$, K.~Ishihara$^5$, A.~Suzuki$^5$, \\
\vspace*{0.1in}
A. Konaka$^6$ \\
\footnotesize
$^1$ Institute for Cosmic Ray Research, University of Tokyo, Kashiwa, Chiba 277-8582, Japan\\
\footnotesize
$^2$ Inst. of Particle and Nuclear Studies, High Energy Accelerator Research Org. 
(KEK), \\ 
\footnotesize
Tsukuba, Ibaraki 305-0801, Japan\\
\footnotesize
$^3$ Department of Physics, Kobe University, Kobe, Hyogo 657-8501, Japan\\
\footnotesize
$^4$ Department of Physics, Kyoto University, Kyoto 606-8502, Japan \\
\footnotesize
$^5$ Department of Physics, Tohoku University, Sendai, Miyagi, 980-8578, Japan \\
\footnotesize
$^6$ TRIUMF, 4004 Wesbrook Mall, Vancouver, British Columbia, Canada, V6T 2A3 \\
}
\date{}
\maketitle
\abstract{
The JHF-Kamioka neutrino project is a second generation long base 
line neutrino oscillation experiment that probes physics beyond the Standard
Model by high precision measurements of the neutrino masses and mixing.
A high intensity narrow band neutrino beam is produced 
by secondary pions created by a high intensity
proton synchrotron at JHF (JAERI).
The neutrino energy is tuned to the oscillation maximum 
at $\sim$1~GeV for a baseline length of 295~km towards the
world largest water \v{C}erenkov detector, Super-Kamiokande.
Its excellent energy resolution and particle
identification enable the reconstruction of the initial
neutrino energy, which is compared with the narrow band neutrino
energy, through the quasi-elastic interaction.
The physics goal of the first phase is 
an order of magnitude better precision
in the $\nu_\mu\rightarrow \nu_\tau$ oscillation measurement
($\delta(\Delta m_{23}^2)=10^{-4}$~eV$^2$ 
and $\delta(\sin^22\theta_{23})=0.01$),
a factor of 20 more sensitive search in the $\nu_\mu\rightarrow\nu_e$
appearance ($\sin^22\theta_{\mu e}\simeq 0.5\sin^22\theta_{13}>0.003$),
and a confirmation of the $\nu_\mu\rightarrow \nu_\tau$ oscillation
or discovery of sterile neutrinos by detecting the neutral current
events.
In the second phase, 
an upgrade of the accelerator from 0.75~MW to 4~MW in  
beam power and the construction of 1~Mt Hyper-Kamiokande 
detector at Kamioka site are envisaged. 
Another order of magnitude improvement in the $\nu_{\mu}\rightarrow\nu_e$
oscillation sensitivity,
a sensitive search of the CP violation in the lepton sector 
(CP phase $\delta$ down to $10^\circ-20^\circ$), 
and an order of magnitude improvement in the proton decay sensitivity 
is also expected.}

\newpage
\normalsize
\tableofcontents
\newpage
\normalsize

\section{Introduction}
\subsection{Physics Motivation}
\hspace*{\parindent}
The developments of high-energy physics and astrophysics in 
twentieth century 
have finally reached the stage where the universe can be described 
upto $\sim10^{-11}$ second after its birth
when electro-weak phase transition took place.
However, many mysteries remain to be explored. 
One of the most profound one is the asymmetry of particles and 
anti-particles in the universe. 
Another one is the generation problem; i.e. why three
generations of quarks and leptons exist,
how are their masses and mixing are determined, and 
whether quarks and leptons are unified at ultra-high energy. 
All these questions may closely be related, but they are out of scope of the 
Standard Model of high energy physics. 
The JHF-Kamioka neutrino project aims at the studies of physics beyond the 
Standard Model through precision measurements of the masses and mixing of 
leptons which seem to be very different from those of quarks.

The discovery of the existence of neutrino oscillation in the atmospheric 
neutrinos by Super-Kamiokande ~\cite{ATM} has opened the possibility of detailed 
studies of the masses and mixing in the lepton sector. 
The first accelerator-based long baseline neutrino oscillation experiment, 
K2K, has started~\cite{k2k}.
Its first result already show the indication of the 
oscillation. 
The establishment of neutrino oscillation in the first generation 
experiment should be followed by high precision measurements of 
neutrino oscillation with much more powerful accelerator.

\subsection{Neutrino Oscillation} \label{nuosc}
\hspace*{\parindent}
The lepton mixing is described by a unitary 3x3 matrix
(Maki-Nakagawa-Sakata~\cite{MNS} (MNS) matrix) that is
defined by a product of three rotation matrices with three
angles ($\theta_{12}$, $\theta_{23}$, and  $\theta_{13}$)
and complex phase ($\delta$) as in Cabibbo-Kobayashi-Maskawa
matrix~\cite{CKM}.
\begin{equation}
\left(
      \begin{array}{@{\,}c@{\,}}
       \nu_e \\
       \nu_{\mu} \\
       \nu_{\tau}
      \end{array}
\right)
      \begin{array}{@{\,}c@{\,}}
       = 
      \end{array}
\left[
      \begin{array}{@{\,}c@{\,}}
     \ \\
  U_{\alpha i}\\
     \  
      \end{array}
\right]
\left(
      \begin{array}{@{\,}c@{\,}}
       \nu_1 \\
       \nu_2 \\
       \nu_3
      \end{array}
\right),
\label{eqn:mixing}
\end{equation}
%
\begin{equation}
U = 
 \left(
     \begin{array}{ccc}
       1 &  0      & 0 \\
       0 &  C_{23} & S_{23} \\
       0 & -S_{23} & C_{23}
      \end{array}
\right)
 \left(
     \begin{array}{ccc}
       C_{13}& 0 & S_{13}e^{-i\delta}  \\
       0     & 1 & 0 \\
       -S_{13}e^{i\delta}& 0 & C_{13}
      \end{array}
\right)
 \left(
     \begin{array}{ccc}
       C_{12}&S_{12}&0\\
       -S_{12}&C_{12}&0 \\
       0&0&1
      \end{array}
\right),
\label{eqn:matrix}
\end{equation}
%
where
$\alpha=e,\ \mu,\ \tau$ are the flavor indices, $i$=1, 2, 3 are the 
indices of the mass eigenstates,  $S_{ij}$ ($C_{ij}$) stands
for $\sin\theta_{ij}$ ($\cos\theta_{ij}$). Neutrinos are
produced as flavor eigenstates and each component of mass eigenstates
gets a different phase after traveling a certain distance. The
detection of neutrino by charged current interactions
projects these new state back onto flavor eigenstates. The
probability of oscillation is given by the formula,
%
\begin{eqnarray}
P(\nu_{\alpha}\rightarrow \nu_{\beta})=\delta_{\alpha\beta}
&  -  & 4\sum_{i>j}\text{Re}(U_{\alpha i}^*U_{\beta i}U_{\alpha j}U_{\beta j}^*)
        \cdot\sin^2\!\Phi_{ij}\nonumber \\
& \pm & 2\sum_{i>j}\text{Im}(U_{\alpha i}^* U_{\beta i} U_{\alpha j} U_{\beta j}^*)
        \cdot\sin\!2\Phi_{ij} \label{eqn:Posc_gen}
\end{eqnarray}
\vspace{-3mm}
where
\begin{equation}
\Phi_{ij}\equiv \Delta m^2_{ij}L/4E_{\nu} =
1.27\Delta m^2_{ij}[\text{eV}^2]L[\text{km}]/E_\nu[\text{GeV}],
\label{eqn:phase}
\end{equation}
$\Delta m^2_{ij}=m^2_j-m^2_i$, $L$ is the flight
distance, and $E_{\nu}$ is the neutrino energy. The $\pm$
sign in the third term is the CP violation effect, $-$ for
neutrinos and $+$ for anti-neutrinos.
Because $\Delta m^2_{12}+\Delta m^2_{23}+\Delta m^2_{31} = 0$,
there exist only two independent $\Delta m^2$ for
three species of neutrinos. Thus 3 generation neutrino oscillation can be
described by two $\Delta m^2$'s, three angles ($\theta_{12},
\theta_{23}, \theta_{13}$) and one phase ($\delta$).

We take the two $\Delta m^2$ values as the values
suggested by solar and atmospheric neutrino measurements;
$\Delta m^2_{12}\equiv\Delta
m^2_{\text{sol}}\simeq10^{-4}\!\to\!10^{-10}$~eV$^2$ and
$\Delta m^2_{23}\simeq\Delta
m^2_{31}\equiv\Delta m^2_{\text{atm}}=(1.6\sim4)\times 10^{-3}$~eV$^2$.
For an oscillation measurement with $E_\nu\simeq\Delta
m^2_{23}\cdot L$ as in this proposed experiment, the
contribution of $\Delta m^2_{12}$ term is small and the
oscillation probabilities can be approximately expressed by
two mixing angles;
%
\begin{alignat}{3}
P(\nu_{\mu}\rightarrow\nu_e)     &=    & \sin^2\!2\theta_{13}\cdot\sin^2\!\theta_{23}\cdot&\sin^2\!\Phi_{23},\label{eqn:Pme}\\
P(\nu_{\mu}\rightarrow\nu_{\mu}) &= 1\ -\ & \sin^2\!2\theta_{23}\cdot\cos^4\!\theta_{13}\cdot&\sin^2\!\Phi_{23}-P(\nu_{\mu}\rightarrow\nu_e),\label{eqn:Pmm}\\
P(\nu_e\rightarrow\nu_e)         &= 1\ -\ & \sin^2\!2\theta_{13}\quad\cdot\quad  &\sin^2\!\Phi_{23}.\label{eqn:Pee}
\end{alignat}
If we define effective mixing angles as
$\sin^2\!2\theta_{\mu e}\!\equiv
\sin^2\!2\theta_{13}\cdot\sin^2\!\theta_{23}$ and
$\sin^2\!2\theta_{\mu\tau}\!\equiv
\sin^2\!2\theta_{23}\cdot\cos^4\!\theta_{13}$, then the
expressions reduce to the ones in the two flavor approximation;
%
\begin{alignat}{3}
P(\nu_{\mu}\rightarrow\nu_e)     &=       & \sin^2\!2\theta_{\mu e}\cdot&\sin^2\!\Phi_{23},\label{eqn:Pme2}\\
P(\nu_{\mu}\rightarrow\nu_{\mu}) &= 1\ -\ & \sin^2\!2\theta_{\mu\tau}\cdot&\sin^2\!\Phi_{23}-P(\nu_{\mu}\rightarrow\nu_e).\label{eqn:Pmm2}
\end{alignat}

Experimental constraints obtained from $\nu_\mu$
disappearance in the atmospheric neutrino are
$\sin^22\theta_{\mu\tau} > 0.89$ and $1.6\times 10^{-3}<
\Delta m^2_{23} <4\times 10^{-3}$~eV$^2$~\cite{SK1st}.
Solar neutrino observations allow
several allowed regions in oscillation parameter space. 
But even for the largest $\Delta m^2_{\text{sol}}$
solution (Large mixing angle solution: LMA), $\Delta
m^2_{\text{sol}} \lesssim 1.6\times
10^{-4}$~eV$^2$ at 95\% CL~\cite{sksolar}.
The most stringent constraint on $\theta_{13}$ comes from
reactor $\bar{\nu}_e$ disappearance experiments. As can be
seen in eq.~(\ref{eqn:Pee}), $\bar{\nu}_e$ disappearance directly
measures $\theta_{13}$. The current limit is
$\sin^22\theta_{13} < 0.05$ for $\Delta m^2_{\text{23}} \simeq 6\times
10^{-3}$~eV$^2$ and $\sin^22\theta_{13} \lesssim 0.12$
for $\Delta m^2_{\text{23}}
\simeq 3\times 10^{-3}$~eV$^2$ at 90 \% C.L~\cite{tht13}.
Since $\theta_{13}$ is very small and atmospheric neutrino data
indicates almost full mixing $\theta_{23}\simeq\pi/4$,
the effective two flavor mixing angles defined above and 3
flavor angles have following approximate relations;
\begin{equation}
\sin^22\theta_{\mu\tau} \simeq
\sin^22\theta_{23},\qquad  \sin^22\theta_{\mu e} \simeq
\frac{1}{2}\sin^22\theta_{13} \simeq 2\left|U_{e3}\right|^2. \label{int:eq:theta}
\end{equation}

It follows from eq.~(\ref{eqn:Posc_gen}) that CP violation
can be observed only with appearance experiments, since
$\text{Im}(U_{\alpha i}^* U_{\beta i} U_{\alpha j} U_{\beta
j}^*)=0$ for $\alpha = \beta$.
Especially $\nu_\mu
\leftrightarrow \nu_e$ oscillation is known to provide the
best chance in measuring the CP asymmetry in lepton
sector. This is because the leading CP
conserving term of $\nu_\mu \leftrightarrow \nu_e$ oscillation is highly
suppressed due to small $\Delta m^2_{12}$ and the subleading
terms, such as $U_{e3}$ related and CP violating terms, give
leading contributions, as shown below.
If the oscillation $\nu_{\mu}\rightarrow\nu_e$ 
is at the observable level in the first phase of the project, further 
investigation of CP violation will be carried out in the second phase.
In addition
since CP violation in three generations requires that all three members should be 
involved in the process, solar neutrino related quantities ($\theta_{12}$, $\Delta_{12}$)
must be large (namely large mixing angle solution for solar neutrino
oscillation) in order for the CP violation of the neutrino oscillation to be observable.

The $\nu_\mu\rightarrow\nu_e$ appearance
probability can be written using MNS matrix element as~\cite{cpformula}
%
\begin{alignat}{1}
 P(\nu_\mu\rightarrow \nu_e) 
 =& 4C^2_{13}S^2_{13}S^2_{23}\sin^2\!\Phi_{31} \nonumber \\
+& 8 C^2_{13}S_{12}S_{13}S_{23}(C_{12}C_{23}\cos\!\delta - S_{12}S_{13}S_{23})
\cos\!\Phi_{32}\cdot\sin\!\Phi_{31}\cdot\sin\!\Phi_{21} \nonumber \\
-& 8 C^2_{13}C_{12}C_{23}S_{12}S_{13}S_{23}\sin\!\delta
\sin\!\Phi_{32}\cdot\sin\!\Phi_{31}\cdot\sin\!\Phi_{21} \nonumber \\
+& 4S^2_{12}C^2_{13}\left(C^2_{12}C^2_{23}+S^2_{12}S^2_{23}S^2_{13}-2C_{12}C_{23}S_{12}S_{23}S_{13}\cos\!\delta\right)
\sin^2\!\Phi_{21}\nonumber \\
-& 8C^2_{13}S^2_{13}S^2_{23}\left(1-2S^2_{13}\right)\frac{aL}{4E_\nu}\cos\!\Phi_{32}\sin\!\Phi_{31}.
\label{eqn:nueapp}
\end{alignat}
%
The first term has the largest contribution. The
second $\cos\!\delta$ term is generated by CP phase $\delta$
but CP conserving. The third $\sin\!\delta$ term violates CP.
The fourth term, which is the solar
neutrino term, is suppressed by $\sin^2\frac{\Delta
m_{21}^2L}{4E_\nu}$.
The matter effect is characterized by
\begin{equation}
a=2\sqrt{2}G_Fn_eE_\nu=7.6\times10^{-5} \rho
[\text{g/cm}^3]E_\nu[\text{GeV}]\qquad [\text{eV}^2],
\label{eqn:matter}
\end{equation}
where $G_F$ is the Fermi constant, $n_e$ is the electron density and
$\rho$ is the earth density.
The probability $P(\bar{\nu}_\mu\rightarrow\bar{\nu}_e)$ is
obtained by the replacing $a\rightarrow -a$ and $\delta
\rightarrow -\delta$ in eq.~(\ref{eqn:nueapp}).
As seen in eq.~(\ref{eqn:matter}) the matter effect is
proportional to neutrino energy, so the lower the energy,
the smaller the effect is.
The CP asymmetry in the absence of matter effect is calculated  as
%
\begin{eqnarray}
A_{CP}&=&
\frac{P(\nu_\mu \rightarrow \nu_e)
  -P(\bar\nu_\mu \rightarrow \bar\nu_e)}{
 P(\nu_\mu \rightarrow \nu_e) 
  +P(\bar\nu_\mu \rightarrow \bar\nu_e)} \simeq
\frac{\Delta m_{12}^2 L}{4E_\nu}
\cdot \frac{\sin\!2\theta_{12}}{\sin\!\theta_{13}} \cdot \sin\!\delta
\end{eqnarray}
Because $\theta_{13}$ is small, CP asymmetry can be large,
especially for small $E_{\nu}$.

\section{Overview of the experiment} \label{jhf}
\begin{figure}[htbp]
\renewcommand{\baselinestretch}{1}
\centerline{\epsfig{figure=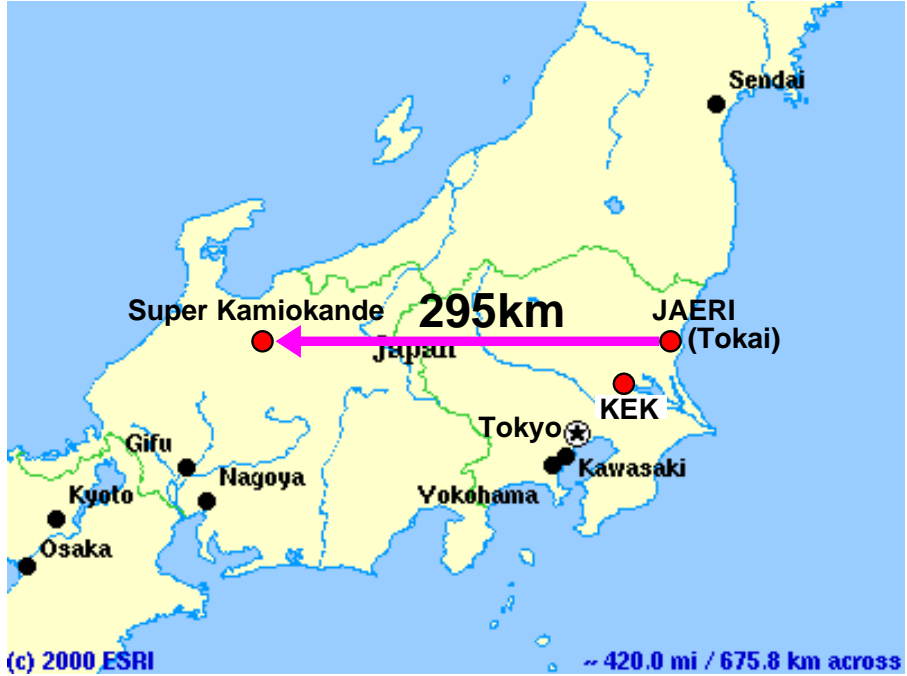,width=12cm}}
\vspace{-3mm}
\caption{\protect\footnotesize
Baseline of the JHF-Kamioka neutrino project}
\label{jhf:fig:map}
\end{figure}

\hspace*{\parindent}
The JHF-Kamioka neutrino project is a long baseline neutrino oscillation experiment using the 
JHF 50~GeV proton synchrotron (PS). The JHF project was officially approved
in 2001 by the Japanese government. The construction of the 50~GeV PS will be completed 
in 2006. The 50~GeV~PS is 
designed to deliver $3.3 \times 10^{14}$ protons every $3.4$ seconds (0.77~MW), 
later to be upgraded to 4~MW~\cite{mori}. A high intensity narrow-band neutrino 
beam of energy around 1~GeV is produced by using the full proton intensity.
The initial far detector is Super-Kamiokande, 
a 50~kt water \v{C}erenkov detector. The baseline length of the 
experiment for neutrino oscillation is 295~km between JHF at Tokai and 
Super-Kamiokande at Kamioka as shown in Figure~\ref{jhf:fig:map}. 
A 1~Mt Water \v{C}erenkov detector, Hyper-Kamiokande to be built at
Kamioka, is proposed as the second far detector. 

A feature of the JHF-Kamioka neutrino experiment is its high intensity narrow band 
neutrino beam. The neutrino energy of narrow band beam is tunable, and
the well-defined energy spectrum has an advantage to achieve the maximum 
sensitivity for neutrino oscillation. The peak of the beam energy will be tuned 
at the point of maximum oscillation and the width will be narrowed to suppress the 
background from non-oscillated neutrinos. With 295~km baseline, the neutrino energy 
will be tuned to between 0.4 and 1.0~$\rm GeV$, which corresponds to $\Delta 
m^2$ between $1.6 \times 10^{-3}$ and $4 \times 10^{-3}$~$\rm eV^2$ suggested by 
the recent Super-Kamiokande result~\cite{SK1st}. In the design of the neutrino beam 
line, a shorter decay volume is adopted to lower the $\nu_e$ flux from muon decays, 
which is an intrinsic background in $\nu_\mu \rightarrow \nu_e$ search.
The narrow band neutrino beam is produced with a conventional method in which 
parent pions are focused by two magnetic horns. There are two methods to make
the narrow band beam: one is momentum-selection of the parent pions 
by a dipole magnet and the other is a technique  called off-axis tuning described in 
Section~\ref{beam}.

\begin{figure}[htbp]
\renewcommand{\baselinestretch}{1}
\centerline{\epsfig{figure=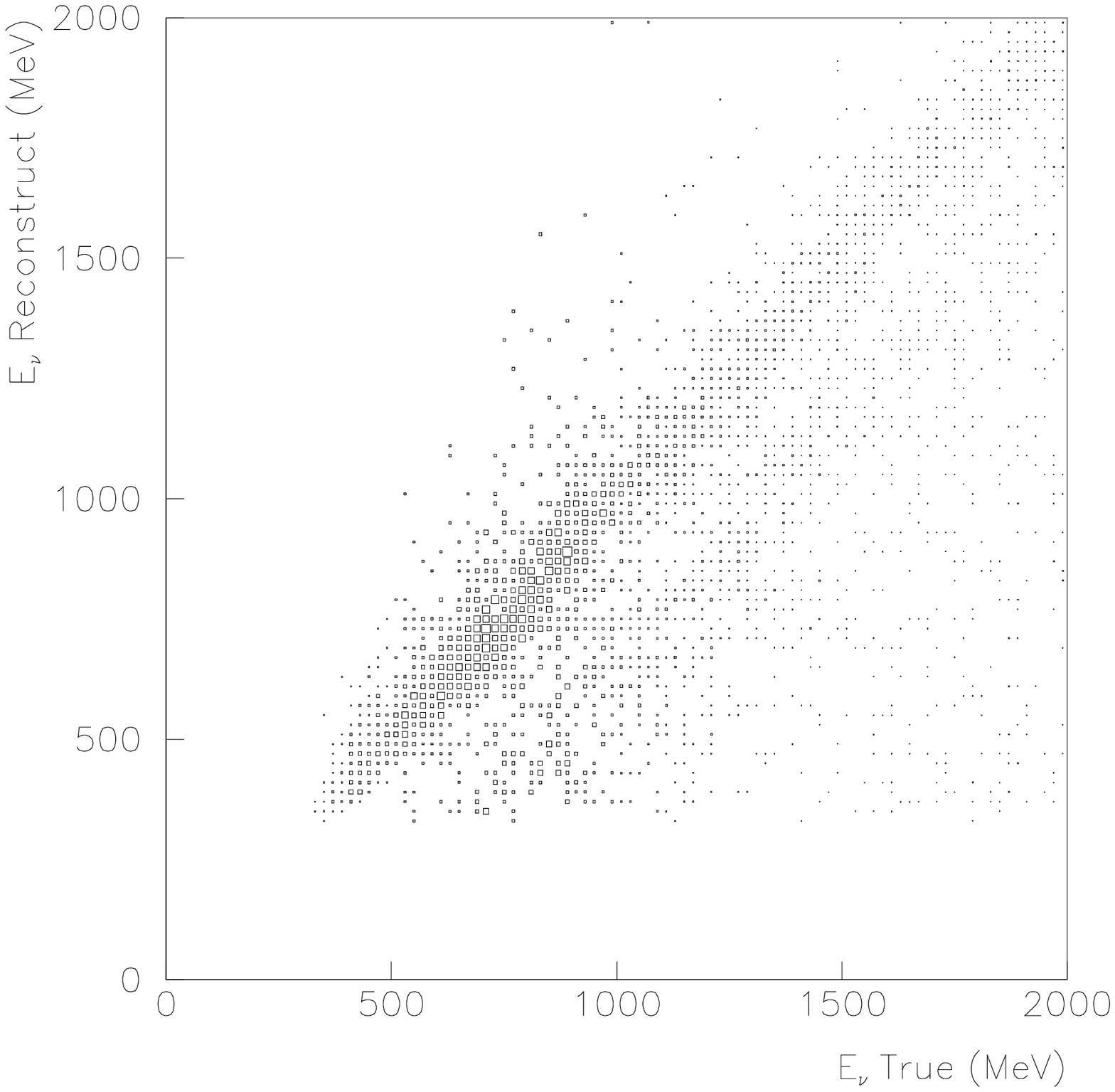,width=8.5cm}
            \epsfig{figure=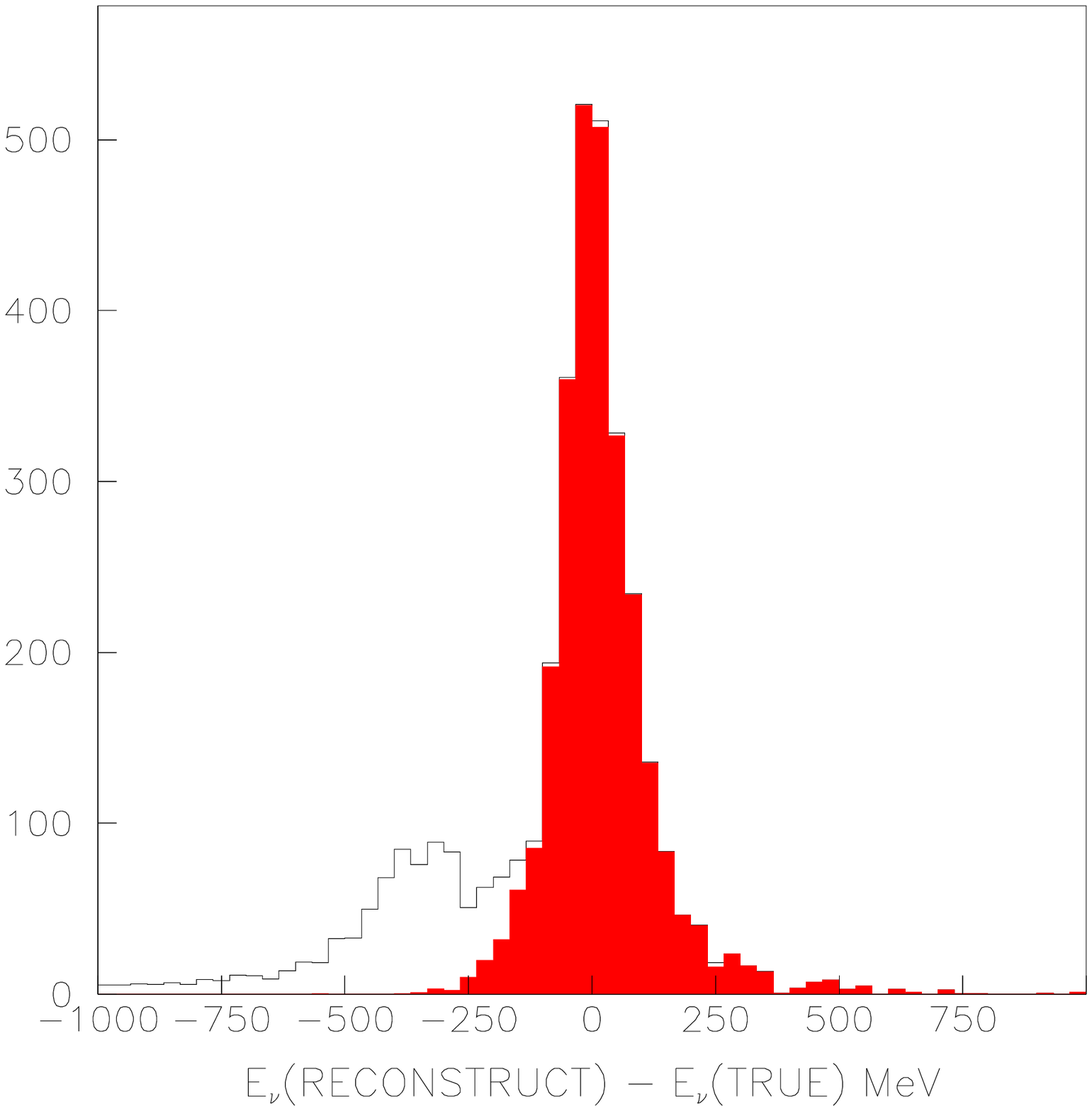,width=8.5cm}}
\vspace{-3mm}
\caption{\protect\footnotesize
(left) The scatter plots of the reconstructed neutrino energy versus
the true one for $\nu_\mu$ events. 
The method of the energy reconstruction is expressed in Equation~\ref{jhf:eq:qe}. 
(right) The energy resolution of $\nu_\mu$ events for 2~degree off-axis beam. 
The shaded (red) histogram is for the true QE events.}
\label{jhf:fig:qe}
\end{figure}

Another feature of the experiment is the reconstruction of the neutrino energy using 
quasi-elastic (QE) interaction. The QE interaction is the dominant process 
at a neutrino energy around 1~$\rm GeV$. In QE interaction, the neutrino 
energy can be well reconstructed with the charged lepton in charged current interaction. 
In QE interaction, the neutrino energy is expressed as:
\begin{eqnarray}
 E_{\nu} & = & \frac{m_N E_l - m_l^2/2}{m_N-E_l+p_l cos \theta_l}, \label{jhf:eq:qe}
\end{eqnarray}
where $m_N$ and $m_l$ is the mass of the neutron and the lepton (=e or $\mu$), and 
$E_l$, $p_l$ and $\theta_l$ is the energy, the momentum, and 
the angle of the lepton relative to the neutrino beam, respectively.
This method matches well with the characteristics of the water \v{C}erenkov detector 
which has an excellent performance for electrons, muons and
photons in this energy scale.
As an example, the relation between the reconstructed neutrino energy and
the true one, and the energy resolution of $\nu_\mu$ events are shown in 
Figure~\ref{jhf:fig:qe} for 2~degree off-axis beam explained in Section~\ref{beam}. 

The project is divided into two phases. 
In the first phase, the main physics goal 
is the precision measurement of neutrino oscillation with the 50~GeV~PS of 
0.77~MW beam power and Super-Kamiokande. 
Typically, $5,000$~$\nu_\mu$ interactions per year 
in the full volume of the detector are expected without neutrino oscillation. 
In $\nu_\mu \rightarrow \nu_\mu$ oscillation channel, 
a measurement of the oscillation pattern determines
$\sin^22\theta_{23}$ with $1 \%$ precision
and $\Delta m^2$ with a precision better than $10^{-4}$~$\rm eV^2$.
A non-oscillation scenario, such as a model of neutrino decay or 
a model with extra dimension~\cite{ed}, can be ruled out in the measurement.
A search for $\nu_\mu \rightarrow \nu_e$ appearance channel is conducted with a 
sensitivity of $\sin^22\theta_{\mu e} (\equiv 0.5 \sin^2 2\theta_{13})$ 
down to $0.003$~$(\rm 90 \% C.L.)$.
Many theories suggest large value of $\sin^22\theta_{\mu e}$~\cite{nue}.
A confirmation of $\nu_\mu \rightarrow \nu_\tau$ oscillation is 
carried out by counting the number of neutral current interaction. 
The measurement also constrains
the existence of a sterile neutrino or
a model of non-oscillation, which is introduced to 
explain both LSND result and solar neutrino results together
with atmospheric neutrino oscillation. 
In the second phase, there are mainly three physics goals:
(1) search for $\sin^22\theta_{e \mu}$ below $5\times 10^{-4}$ level,  
(2) search for CP violation in neutrino oscillation, and
(3) discovery of the proton decay as an evidence of the Grand unification theory (GUT).
In the search for CP violation, the low energy neutrino beam has the advantage of
large CP asymmetry and small matter effect as discussed in Section~\ref{nuosc}.

The JHF-Kamioka neutrino project is a unique experiment in the world. 
The experiment has much more sensitivity for neutrino oscillation than 
other long baseline neutrino experiments (K2K, MINOS, and OPERA) 
due to the high power JHF accelerator and the huge Super-Kamiokande detector. 
The second phase of the experiment is competitive
with a recent proposal of a neutrino factory~\cite{nufact} in
terms of CP violation reach and uses established technologies. 
The JHF-Kamioka neutrino project is expected to start in 2007 at the same 
time of completion of the JHF accelerator.

\section{Neutrino beam at JHF} \label{beam}

\begin{figure}[htbp]
\renewcommand{\baselinestretch}{1}
\centerline{
\epsfig{file=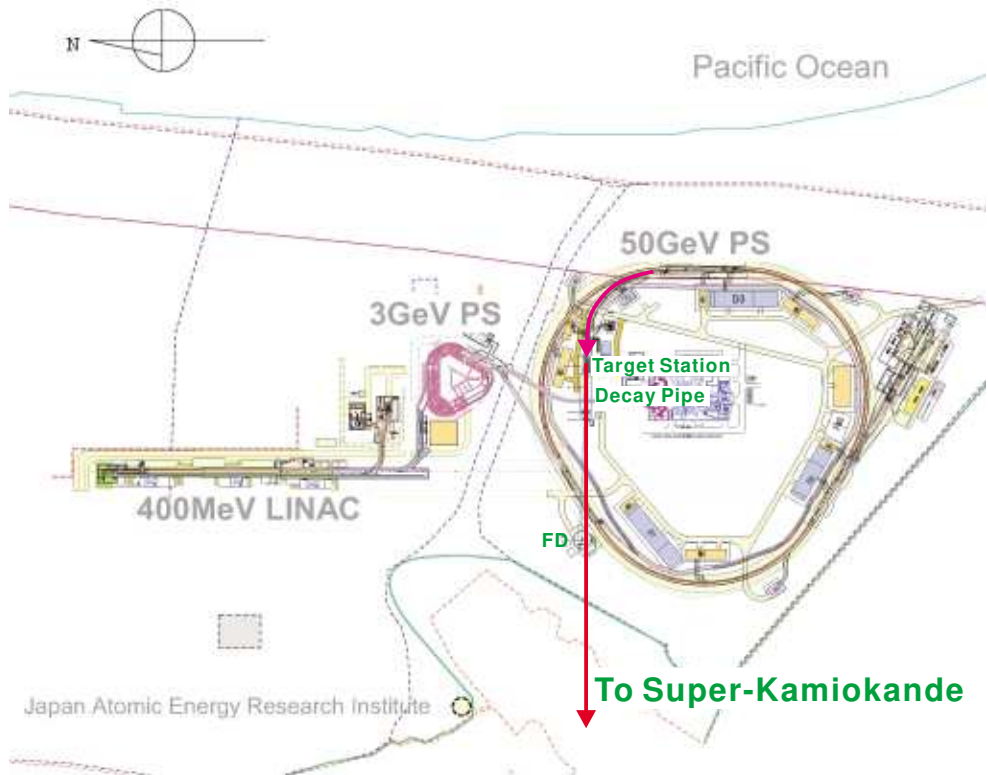,width=13cm}
}
\vspace{-5mm}
\caption{\protect\footnotesize Layout of JHF.}
\label{bmfig:jhf-layout}
\end{figure}

\hspace*{\parindent}
The layout of JHF facility is drawn in Figure~\ref{bmfig:jhf-layout}.
%
%
The proton beam is fast-extracted from the 50~GeV PS in a single
turn and transported to the production target. The design intensity of
the PS is $3.3\times 10^{14}$ protons/pulse (ppp) at a
repetition rate of 0.292~Hz (3.42~second period), resulting in
a beam power of 
0.77~MW (2.64~MJ/pulse). The spill width is $\sim$5.2~$\mu$sec. 
We define a typical one year operation as $10^{21}$ protons on
target (POT), which 
corresponds to about 130 days of operation.
The protons are extracted toward inside of the PS ring, 
and are bent by 90$^\circ$ to SK direction by the transport 
line with a radius of curvature of 110~m smaller than the PS arc.
We will adopt superconducting magnets for the transport line.
The secondary pions (and kaons) from the target are focused by
electromagnetic horns \cite{K2Khorn}, and decay in the decay pipe.
The length of the decay pipe from the target position is 80~m.
The first front detector is located at 280~m from the target.

There are three options of beam configurations;
wide band beam (WBB), narrow band beam (NBB) and off axis
beam (OAB). The difference is in the optics for the
secondary particles.
The WBB uses two electromagnetic horns to
focus secondary pions. Horn-focused WBB has been widely used in
experiments including K2K. Since the
momentum and angular acceptance of the horn system is large,
the resulting neutrino spectrum becomes wide.
The NBB is obtained by placing a dipole magnet between the two
horns of WBB. Momentum selected pion beam produces
a monochromatic neutrino beam. The peak neutrino energy can be 
tuned easily by changing the current of the dipole magnet.
The OAB is another option to produce a narrow neutrino energy
spectrum~\cite{OAB}. The optics is almost same as the WBB, but
the axis of the beam optics is displaced by a few degree from the 
far detector direction (off-axis). With a finite decay angle, 
the neutrino energy becomes almost independent of parent pion 
momentum as a characteristics of the Lorenz boost, which provides the 
narrow spectrum. The peak neutrino energy can be adjusted 
by choosing the off-axis angle.
For each of the above beam configurations, $\nu_\mu$ and
$\bar{\nu}_\mu$ can be switched by flipping the polarity of
the focusing magnets.

Monte-Carlo (MC) simulations using GEANT~\cite{GEANT}
have been performed
to estimate expected neutrino spectra and number of
events. The target, horns, bending magnet and
decay pipe are put into the geometry of the simulation.
The target is assumed to be simple Cu rod of 1-cm diameter and
30-cm long. Hereafter we refer to the NBB with selected pion
momentum of \#~GeV/c as LE\#$\pi$ and OAB with a beam axis
\# degree offset as OA\#$^\circ$.
The simulation code is essentially the one used in the
K2K experiment except for the hadron production model. The Calor-Fluka
model~\cite{CALOR}, which is known to work better at higher energies,
is used in the present simulation while
a dedicated model base on hadron production data taken near 12~GeV is
used in the K2K. From the observation of neutrino events at
front detector of K2K, the K2K beam MC is known to provide
absolute neutrino flux within 20\% error.

\begin{figure}[!tb]
\renewcommand{\baselinestretch}{1}
\centerline{
\epsfig{file=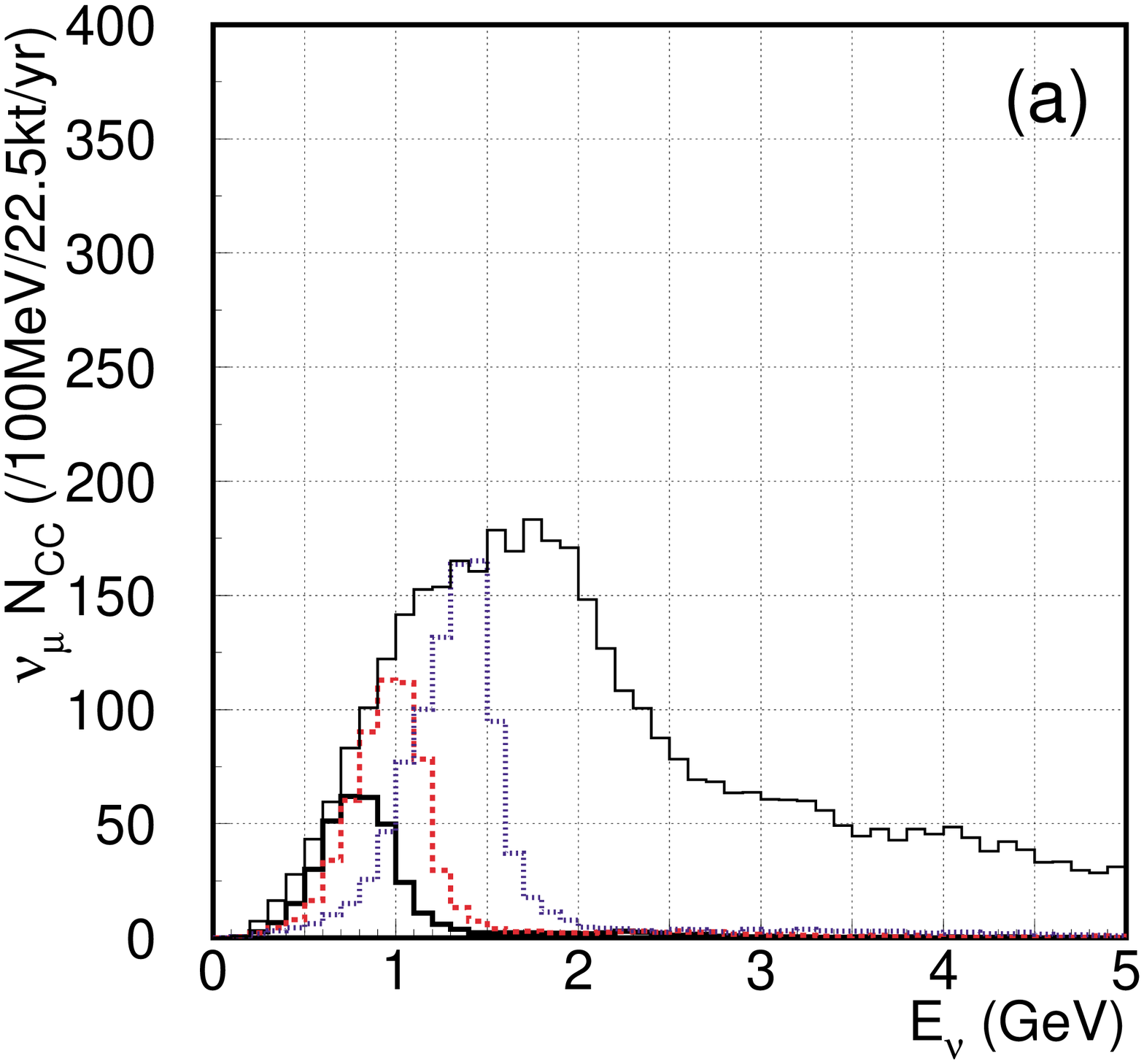,width=7cm}
\hspace{5mm}
\epsfig{file=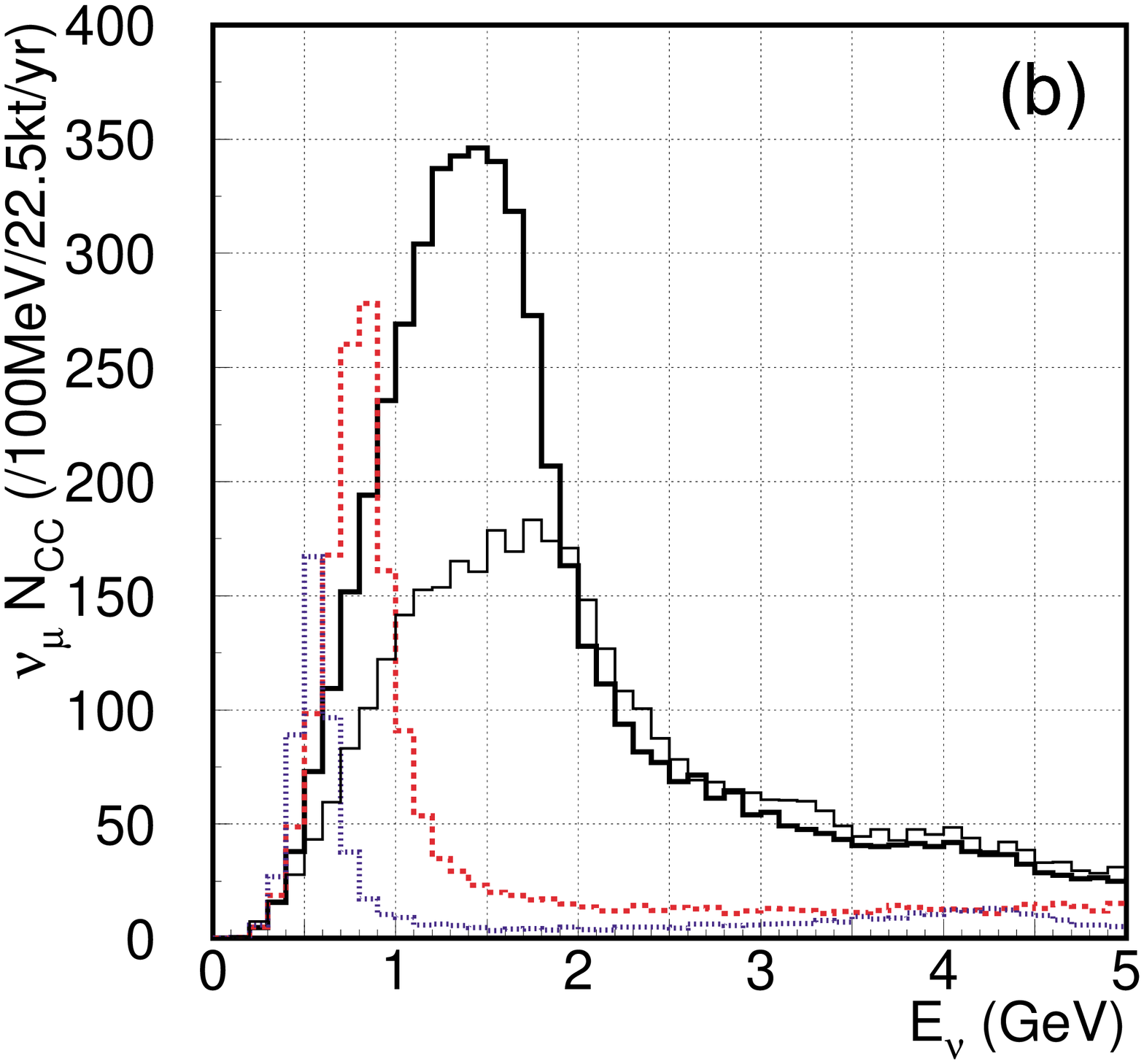,width=7cm}
}
\vspace{-5mm}
\caption{\protect\footnotesize
Neutrino energy spectra of charged current interactions.
Thick solid, dashed and dotted histograms in (a) are
LE1.5$\pi$, LE2$\pi$ and LE3$\pi$, and those in (b) are
OA1$^\circ$, OA2$^\circ$ and OA3$^\circ$, respectively.
WBB is drawn by thin solid histogram in both (a) and (b).}
\label{bmfig:fluxNBWBOA}
\end{figure}

Fig.~\ref{bmfig:fluxNBWBOA} shows expected neutrino energy spectrum
of charged current interactions at Super-Kamiokande.
%
%
Fluxes and numbers of interactions are summarized in
Table~\ref{bmtbl:summary1}.
%
\begin{table}[!tb]
\renewcommand{\baselinestretch}{1}
\caption{\protect\footnotesize Summary of $\nu_\mu$ beam simulation. The peak
energy $E_{\text{peak}}$ is in GeV. The flux is given in
10$^6$/cm$^2$/yr, and the $\nu_e/\nu_\mu$ flux ratio is in
\%. The ratio in the ``total'' column is the one integrated
over neutrino energy and the column ``$E_{\text{peak}}$'' is
the ratio at the peak energy of $\nu_\mu$ spectrum.
The normalization for the number of interactions are
/22.5kt/yr. The numbers outside (inside) the
bracket are number of total (CC) interactions.}
\label{bmtbl:summary1}
\begin{center}
\begin{tabular}{l|l|rr|rr|rr}
\hline\hline
            & & \multicolumn{2}{c}{Flux} & \multicolumn{2}{|c|}{$\nu_e$/$\nu_\mu$(\%)}
            & \multicolumn{2}{c}{\# of interactions}\\

Beam        & \multicolumn{1}{c}{$E_{\text{peak}}$}
            & \multicolumn{1}{|c}{$\nu_\mu$}
            & \multicolumn{1}{c}{$\nu_e$}
            & \multicolumn{1}{|c}{total}
            & \multicolumn{1}{c}{$E_{\text{peak}}$}
            & \multicolumn{1}{|c}{$\nu_\mu$}
            & \multicolumn{1}{c}{$\nu_e$}\\
\hline
WIDE        & 1.1  & 25.5 & 0.19 & 0.74 & 0.34 & 7000(5200) & 78(59)  \\
LE1.5$\pi$  & 0.7  & 5.3 &  0.05 & 1.00 & 0.39 &  510(~360) & 5.7(4.2)\\
LE2$\pi$    & 0.95 & 7.0 &  0.05 & 0.73 & 0.15 &  870(~620) & 6.8(5.0)\\
LE3$\pi$    & 1.4 & 8.0 &   0.05 & 0.65 & 0.16 & 1400(1000) & 9.3(6.9)\\
OA2$^\circ$ & 0.7 & 19.2 &  0.19 & 1.00 & 0.21 & 3100(2200) & 60(45)  \\
OA3$^\circ$ & 0.55 & 10.6 & 0.13 & 1.21 & 0.20 & 1100(~800) & 29(22)  \\
\hline\hline
\end{tabular}
\end{center}
\end{table}
%
LE2$\pi$ and OA2$^\circ$ have a sharp peak at $\sim$0.95~GeV
and $\sim$0.7~GeV, respectively, and WBB has a broad peak
at $\sim$1~GeV.
The OAB is roughly a factor of three more intense than NBB. 
NBB has the least high energy tail among the three beams.
The $\nu_e$ contamination in the beam are expected to be
0.7\%, 0.7\% and 1.0\% for WBB, LE2$\pi$ and OA2$^\circ$. 
The sources of $\nu_e$ are $\pi \rightarrow \mu \rightarrow
e$ decay chain and K decay (K$_{e3}$). Their fractions are
$\mu$-dacay: 54\%, K-decay: 46\% for LE2$\pi$ and 
$\mu$-dacay: 37\%, K-decay: 63\% for OA2$^\circ$.
The energy spectra of the $\nu_e$ contamination are plotted in
Fig.~\ref{bmfig:nuenumu}.
%
\begin{figure}[!tb]
\renewcommand{\baselinestretch}{1}
\centerline{
\epsfig{file=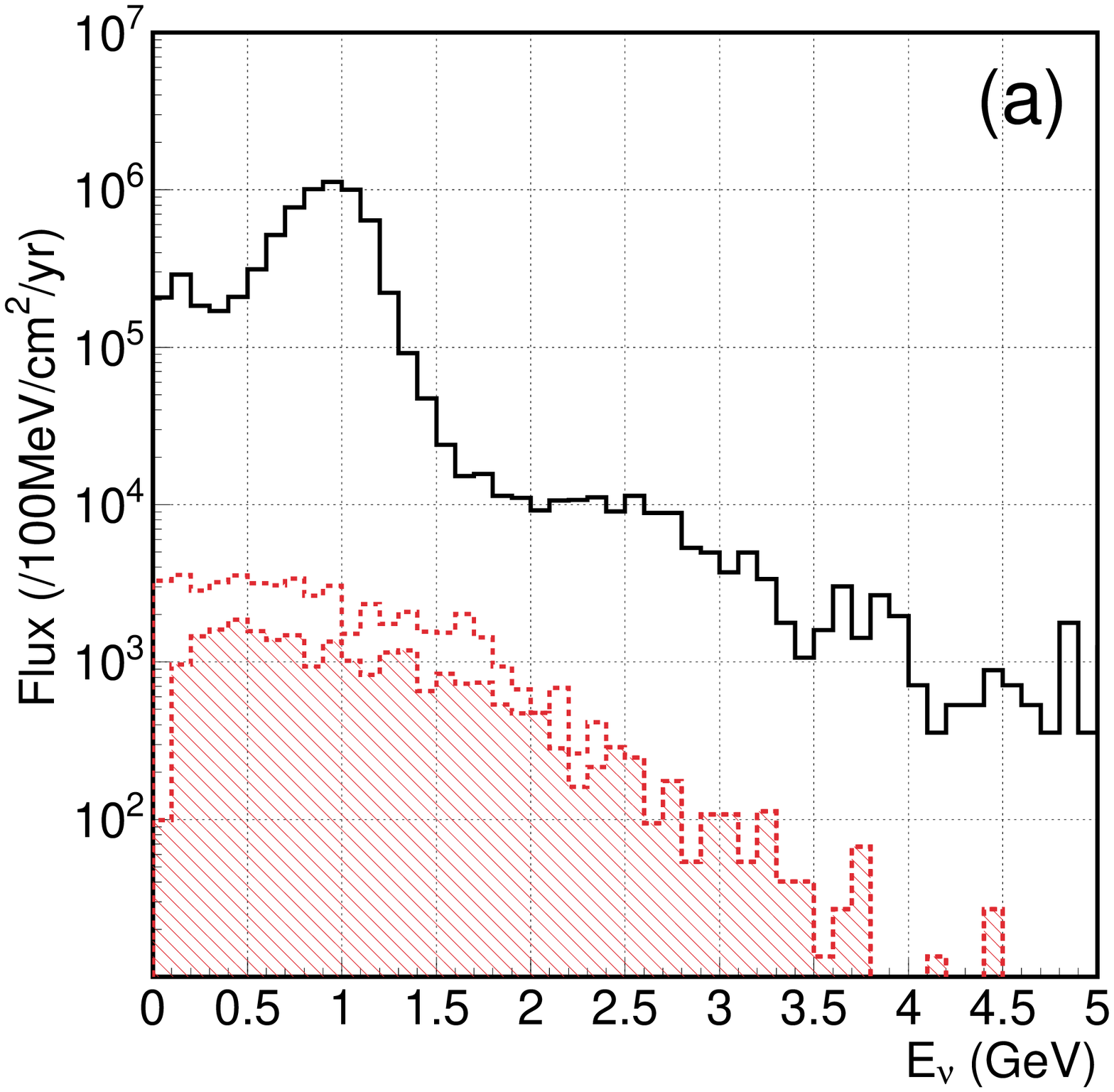,width=7cm}
\hspace{5mm}
\epsfig{file=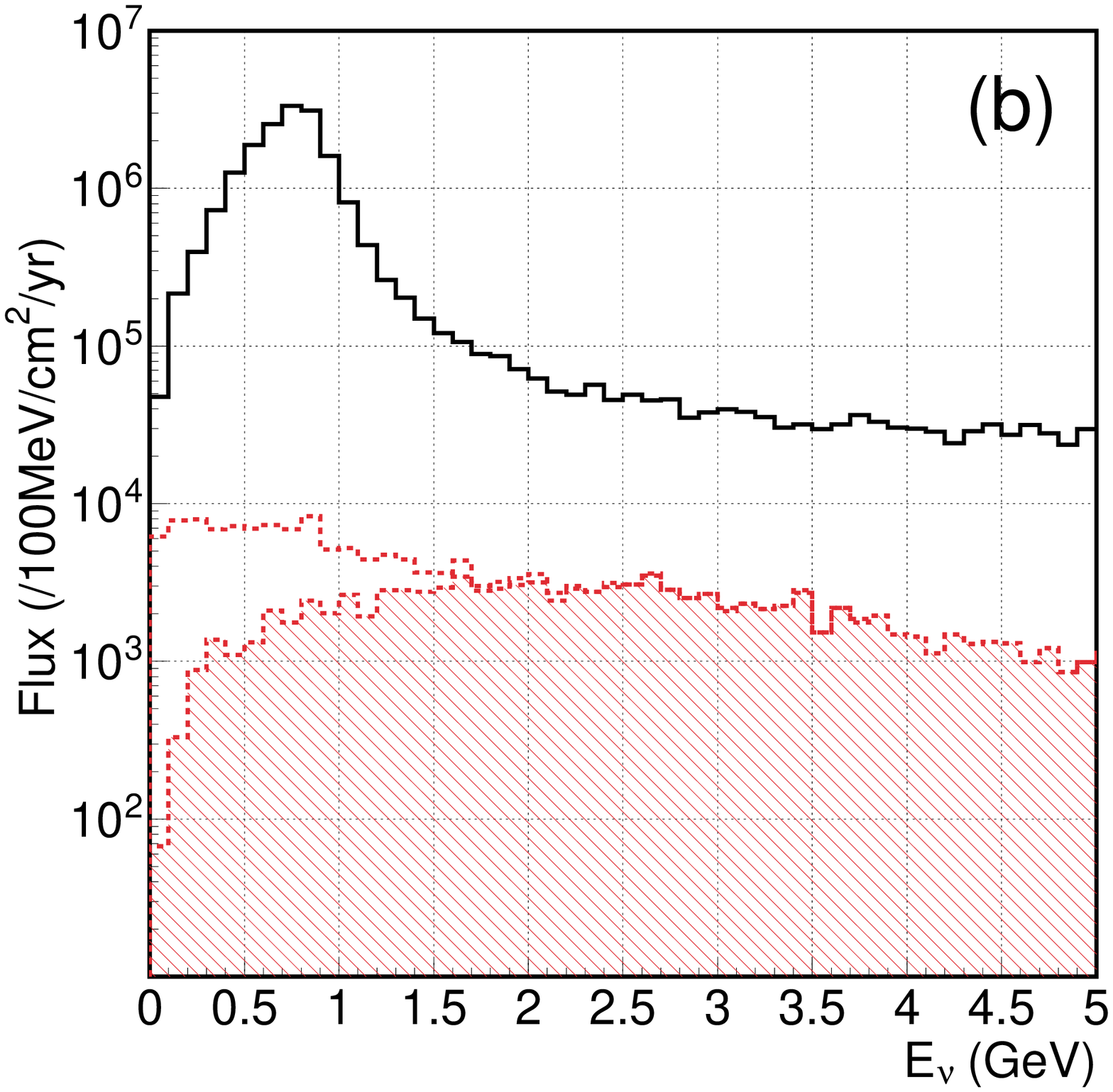,width=7cm}
}
\vspace{-5mm}
\caption{\protect\footnotesize Comparison of $\nu_e$ and $\nu_\mu$ spectra for (a)
LE2$\pi$ and (b) OA2$^\circ$.
Solid (black) histogram is $\nu_\mu$ and dashed (red) one is
$\nu_e$. Hatched area is contribution from K decay.}
\label{bmfig:nuenumu}
\end{figure}
%
At the peak energy of $\nu_\mu$
spectrum, the $\nu_\mu / \nu_e$ ratio is as small as 
0.2\% in the cases of NBB and OAB. This indicates that beam $\nu_e$
background is greatly suppressed (factor $\sim 4$) by
appling energy window cut in the event selection.
The $\mu$ decay is the largest contribution 
at the peak neutrino energy as shown in the figure.

\begin{figure}
\renewcommand{\baselinestretch}{1}
\centerline{
\epsfig{file=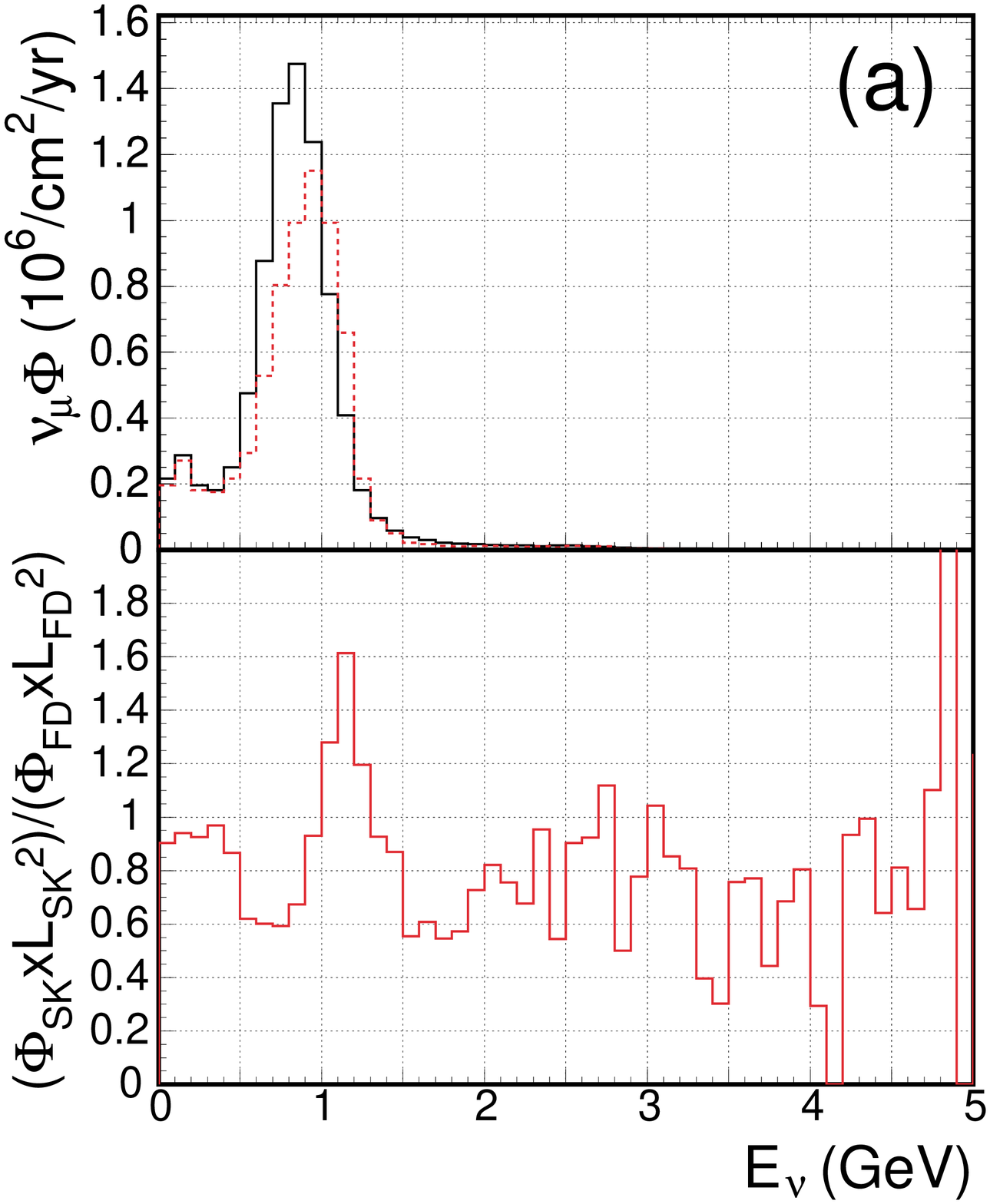,width=5cm}
\epsfig{file=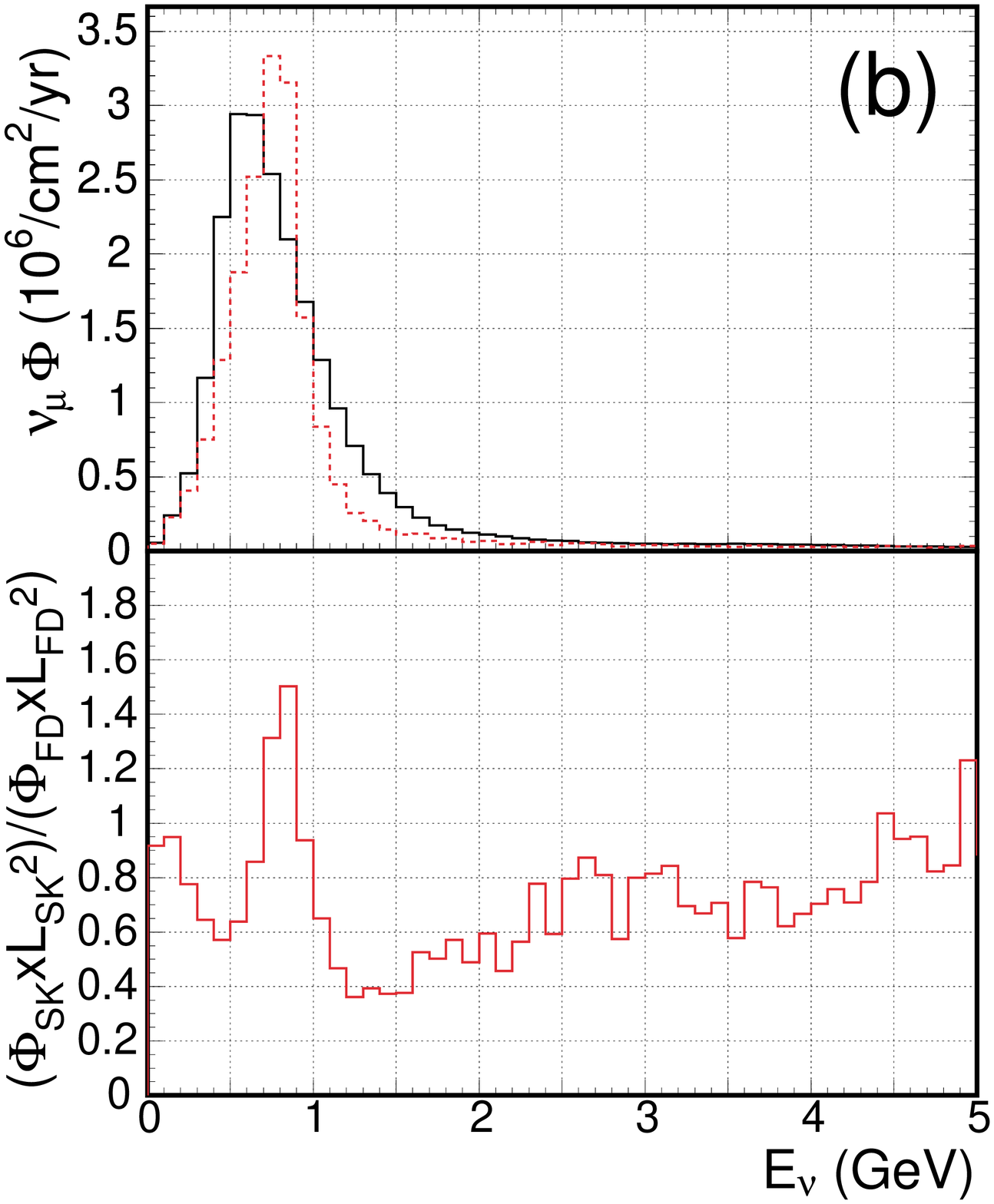,width=5cm}
\epsfig{file=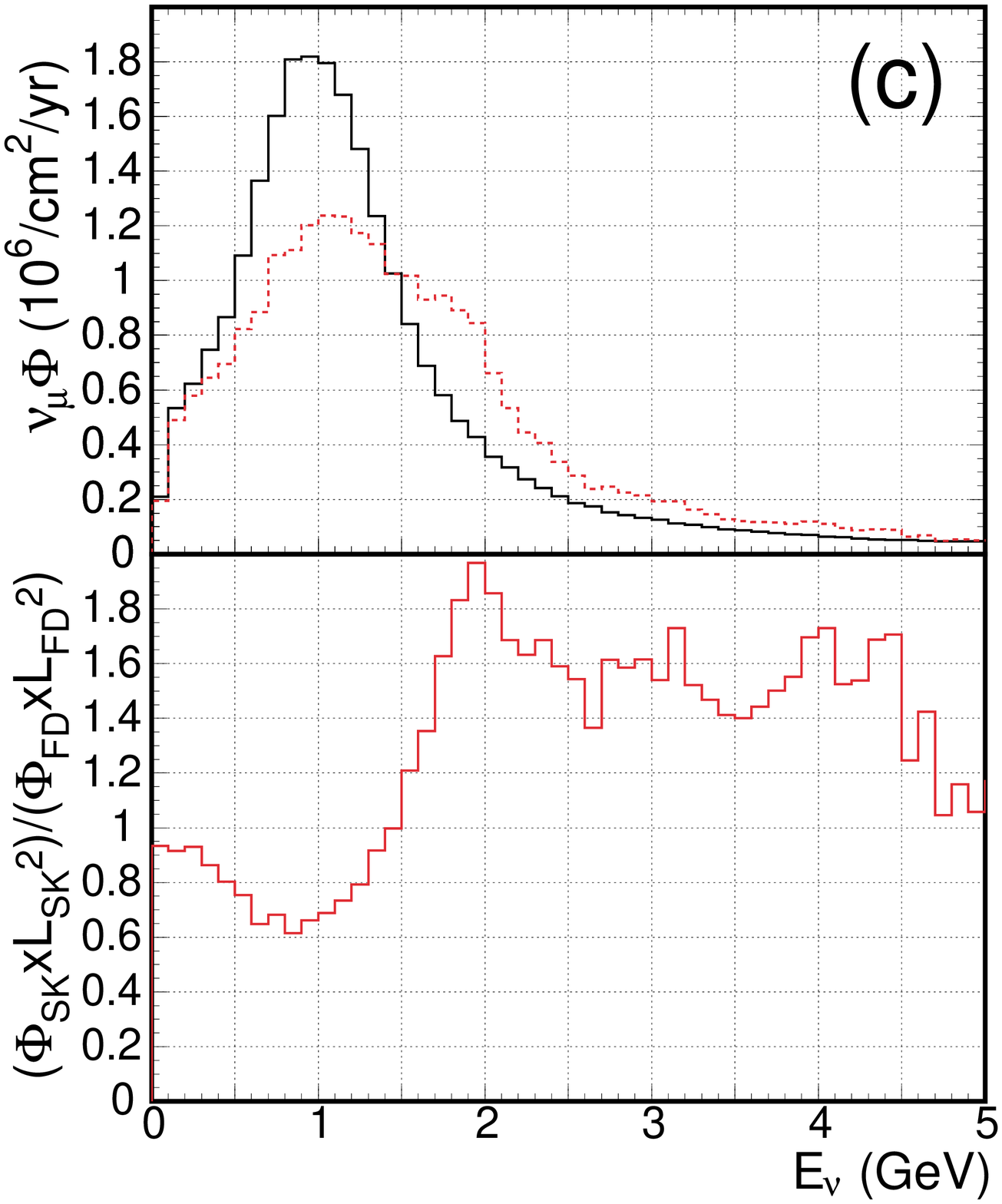,width=5cm}
}
\vspace{-5mm}
\caption{\protect\footnotesize Comparison of spectra at far
and near site for (a) LE2$\pi$, (b) OA2$^\circ$ and (c) WBB.
Upper figure is $\nu_\mu$ spectra at 280~m (solid black
histogram) and 295~km (dashed red histogram). The flux for
the near site is multiplied by $(295/0.28)^2$ to directly
compare the spectra. The front detector size is assumed to
be $\pm 5$~m in horizontal and vertical directions.
The lower plots are far/near ratio of
fluxes.}
\label{bmfig:farnear}
\end{figure}

One of the sources of systematic uncertainties in estimating
the expected number of events in the far detector from the
observation in the near detector is the spectrum
difference between far site and near site.
In Figure~\ref{bmfig:farnear}, spectra at far and near sites
are compared. The peak postion is shifted to higher energy at
the far site than at the near site. The far/near spectrum ratio
is also plotted in the figure. The difference is as
large as 40$\sim$50\% for NBB and OAB and reaches almost 80\%
for WBB.
%
%
The sources of this far/near difference are
\begin{itemize}
\item Difference in the solid angle between far and near
detectors. Larger angle neutrinos from beam axis can
contribute more to the near detector than to the far detector.
The Large angle neutrinos have different energies
from neutrinos at zero degree direction.
\item Finite length of decay pipe. The near
detector has a larger solid angle for pions which decay
near the end of the decay pipe than those decaying at the beginning
of the decay pipe. Higher momentum pions decay further
downstream. For the far detector,
the length of the decay pipe can be neglected as a point source. 
Thus the neutrino spectrum is also distorted by this finite 
decay pipe effect.
\end{itemize}
At the distance longer than one~km from the target, 
both of the above two effects become negligible and 
the far/near ratio becomes flat.

\section{Secondary beam monitor and Neutrino detectors} \label{det}
\subsection{Muon monitor}
\hspace*{\parindent}
Muons from pion decays are measured by the muon monitor. Only
such a beam monitor can monitor the beam condition spill by spill, since
a neutrino detector cannot monitor the beam condition in
such a short period of spill. The primary purpose of the
muon monitor is to measure the beam direction with a 
precision of 1~$\rm mrad$ similar to K2K~\cite{k2k} by measuring the
center of the muon beam profile. The stability of the neutrino
beam intensity is also monitored by measuring the time
dependent muon yield.

The beam monitor is located right after the beam dump which stops all
hadronic activities. We prepare separate beam monitors for each beam,   
since the beam center and the muon energy of NBB is different from
those of OAB. 

\subsection{Near neutrino detector}
\hspace*{\parindent}
The main purposes of the front-detector at the near site are as follows.
\begin{enumerate}
\item Measurement of the neutrino flux to estimate the flux at SK.
\item Measurement of the neutrino energy spectrum.
     The spectrum measurement as a function of the distance from the 
     beam center gives constraint on the estimation of the 
     ``$\rm far/near$'' ratio.
\item Measurement of neutrino cross sections for
      various interaction modes, such as charged current quasi-elastic, 
      charged current non quasi-elastic, and neutral current interactions.
\item Measurement of the $\nu_e$ contamination for the $\nu_e$ appearance
      search.
\item Measurement and monitoring of neutrino beam direction.
\end{enumerate}
 For these purposes, we have several options for the detector.
For low energy neutrinos with energy below 1~GeV, fully active detectors, 
such as water \v{C}erenkov and a fine grained
scintillator detectors satisfy our requirements.
Because typical event rate in the first near detector
is $0.06$~events per spill per ton 
for OAB and $\sim 0.02$~events for NBB, 
it hard to use a water \v{C}erenkov detector
like the 1kt detector of K2K.
The fully active scintillator tracker proposed for  the
K2K upgrade~\cite{k2kup} can be operated at high event rate.
The difference in the neutrino target material between H$_2$O and CH$_2$
needs to be studied using the near detector and the narrow-band neutrino beam.

There is a possibility of a second near detector located
at a few kilometers distance from the target.
At that distance,  
the ``$\rm far/near$ ratio'' is almost flat as discussed in section~\ref{beam}.
In addition, because the neutrino flux is not as high as at the near site, 
a Water \v{C}erenkov detector can be operated with a reasonable 
event rate.
Currently, we look for a place for the detector, and
we study the necessity of the detector at the position.

\subsection{Far detector: Super-Kamiokande }
\begin{figure}[!tb]
\centerline{\psfig{file=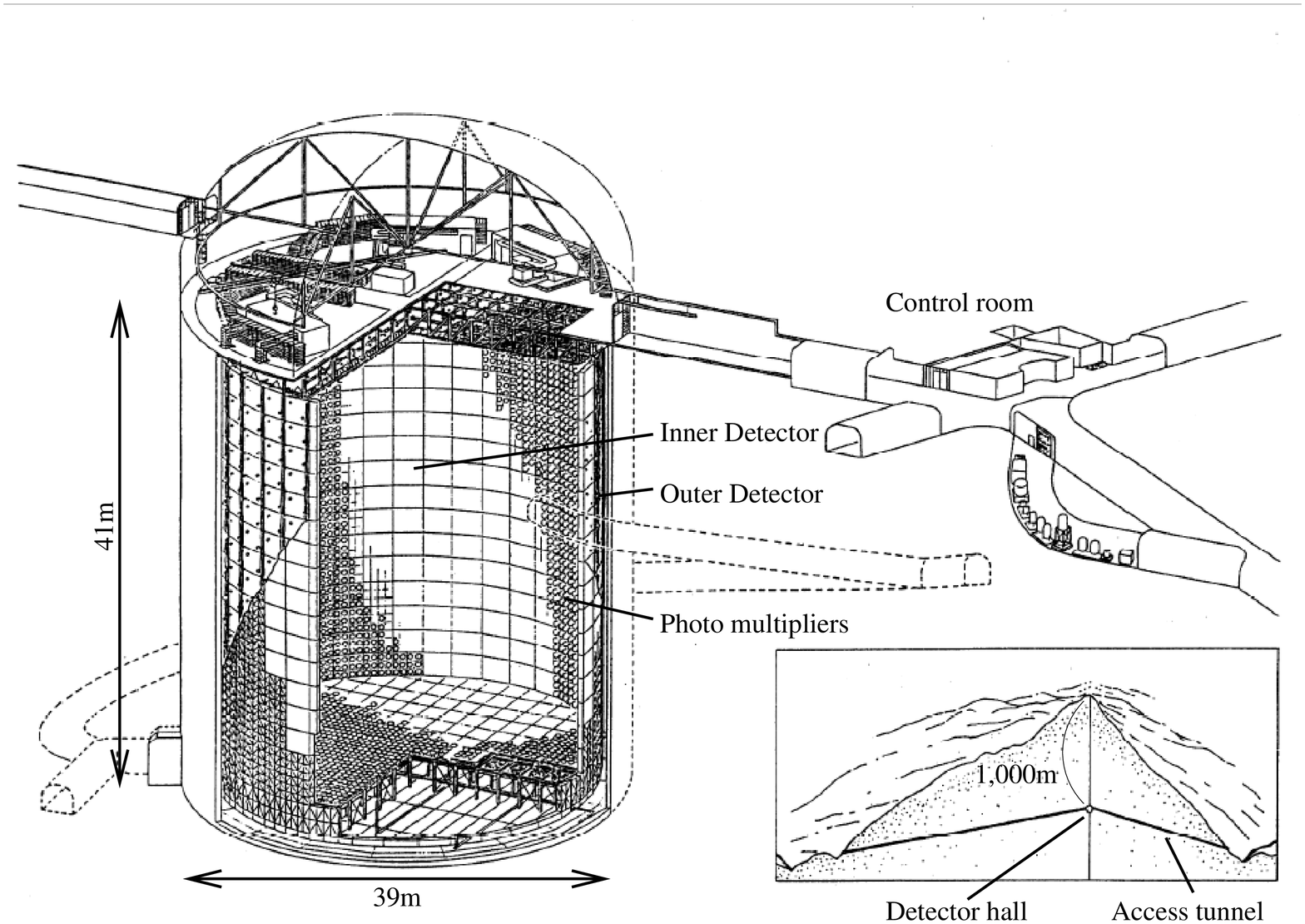,width=10cm}}
\vspace{-4mm}
\caption{\protect\footnotesize
A schematic view of the Super-Kamiokande Detector.}
\label{superk}
\end{figure}

\hspace*{\parindent}
The far detector, Super-Kamiokande, is located in the Kamioka Observatory, 
Institute for Cosmic Ray Research (ICRR), University of Tokyo, 
which has been successfully taking data since 1996.
The detector is also used as a far detector for K2K experiment. 
It is a 50,000~ton water \v{C}erenkov detector located 
at a depth of 2,700~meters water equivalent in the Kamioka mine in Japan. 
Its performance and results in atmospheric neutrinos or solar neutrinos 
have been well documented elsewhere\cite{ATM,SK1st,sksolar}.
A schematic view of detector is shown as Fig~\ref{superk}. 
The detector cavity is 42 m in height and 39 m in diameter, 
filled with 50,000~tons of pure water.
There is an inner detector (ID), 33.8~m diameter and 36.2~m high, 
surrounded by an outer detector (OD) of approximately 2 m thick. 
The inner detector has 11,146 50~cm $\phi$ photomultiplier tubes (PMTs), 
instrumented on  all surfaces of the inner detector on a 70.7~cm grid spacing. 
The outer detector is instrumented with 1,885 20~cm $\phi$ PMTs
and used as an anti-counter to identify entering/exiting particles to/from ID. 
The fiducial volume is defined as 2 m away from the ID wall,
and the total fiducial mass is 22,500~ton.
\v{C}erenkov rings produced by relativistic charged particles 
are detected by ID PMT's.  
The trigger threshold is recently achieved to be 4.3 MeV.
The pulse hight and timing information of the PMT's are
fitted to reconstruct the vertex, direction, energy,
and particle identification of the \v{C}erenkov rings.
A typical vertex, angular and energy resolution for a
1~GeV $\mu$ is 30~cm, 3$^\circ$ and 3\% for vertex, respectively.
The \v{C}erenkov ring shapes, clear ring for muons and fuzzy
ring for electrons, provides good $e$/$\mu$ identification. 
A typical rejection factor to separate $\mu$'s from $e$'s (or vice versa)
is about 100 for a single \v{C}erenkov ring events at 1~GeV.
The $e$'s and $\mu$'s are further separated by detecting
decay electrons from the $\mu$ decays.
A typical detection efficiency of 
decay electrons from cosmic stopping muons is roughly 80\% which can be 
improved by further analysis.  
A 4$\pi$ coverage around the interaction vertex
provides an efficient $\pi^0$ detection
and $e$/$\pi^0$ separation as discussed in sections 5.2 amd 5.3.

Interactions of neutrinos from the accelerator
are identified by synchronizing the timing between
the beam extraction time at the accelerator and
the trigger time at Super-Kamiokande using the
the Global Positioning System (GPS).
The synchronization accuracy 
of the two sites is demonstrated to be less than $200$~ns 
in the K2K experiment.
Because of this stringent time constraint,  
and the quiet environment of the deep Kamioka mine, 
chance coincidence of any entering background is negligibly low.
A typical chance coincidence rate of atmospheric neutrino events is 
$10^{-10}$/spill, which is much smaller than the signal rate of
about 3$\times 10^{-3}$/spill for the WBB option.

\section{Physics in the first stage of the project}

\subsection{High precision measurement of $\Delta m_{23}$ and 
$\theta_{23}$ with $\nu_{\mu}$ disappearance}
\begin{figure}[!tb]
\renewcommand{\baselinestretch}{1}
\centerline{\epsfig{figure=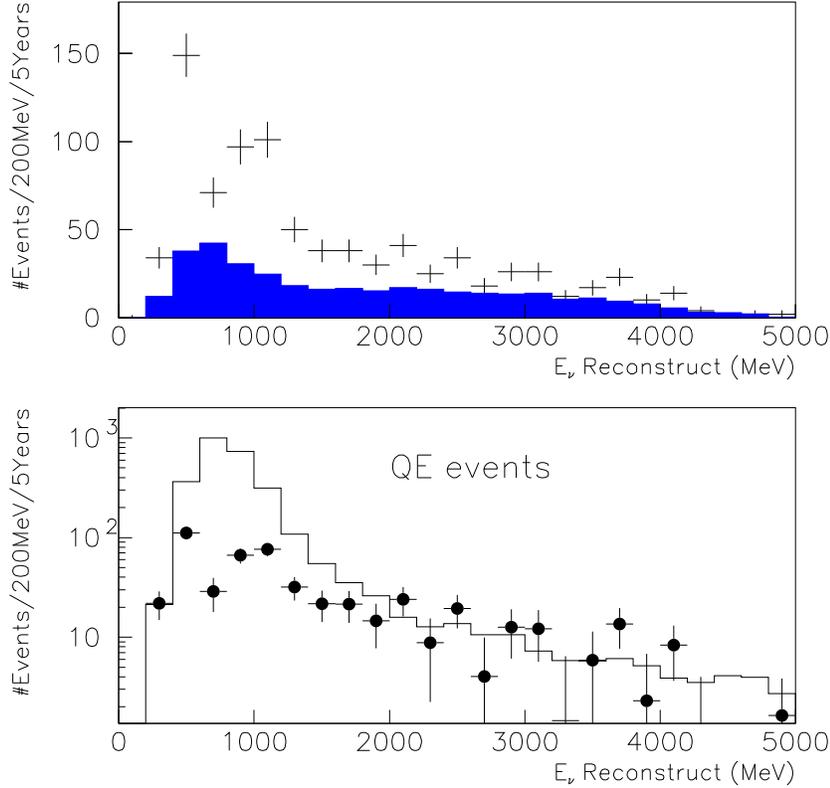,width=12cm}}
\vspace{-5mm}
\caption{\protect\footnotesize
The neutrino energy spectrum measured in SK for 5 
years exposure of JHF OA2$^\circ$.
The top plot is in the case of neutrino oscillation with 
parameters of $(\Delta m_{23}, \theta_{23}) = 
(3\times 10^{-3}$~eV$^2, \pi/4)$.
The contribution of non-QE interactions is shown 
by the shaded (blue) histogram. The bottom plot is for QE events 
after subtracting the non-QE contributions. The dot is the case
of the oscillation corresponding to the top plot, and the 
histogram is in the case of null oscillation. The error bar is 
from the statistics of 5 years.}
\label{fig:numu:ene}
\end{figure}

\hspace*{\parindent}
The neutrino energy can be reconstructed
through quasi-elastic (QE) interactions as shown in Equation~\ref{jhf:eq:qe}
by the Super-Kamiokande (SK) detector. 
In this analysis, we use the same muon selection criteria
as those used in the atmospheric neutrino analysis
by the Super Kamiokande collaboration~\cite{ATM};
fully contained single ring muon-like events 
in a fiducial volume of $22.5$~kt. 
The neutrino energy spectrum at SK
in five years exposure of the JHF OA2$^\circ$ neutrino beam
is shown in 
Figure~\ref{fig:numu:ene}~(bottom) assuming no neutrino oscillation. 
The observed energy spectrum for
the neutrino oscillation with the parameters of
$(\Delta m_{23}, \theta_{23}) = \rm (3\times 10^{-3} eV^2, \pi/4)$ 
is shown in 
Figure~\ref{fig:numu:ene}~(top), which shows large dip at around 800~MeV. 
In this analysis, $\theta_{13}$ is
approximated to be zero and thus $\sin^22\theta_{\mu e} = \sin^22\theta_{23}$. 
The value of $\theta_{13}$ will be determined in the $\nu_\mu \rightarrow
\nu_e$ search. 

\begin{figure}[!tb]
\renewcommand{\baselinestretch}{1}
\centerline{\epsfig{figure=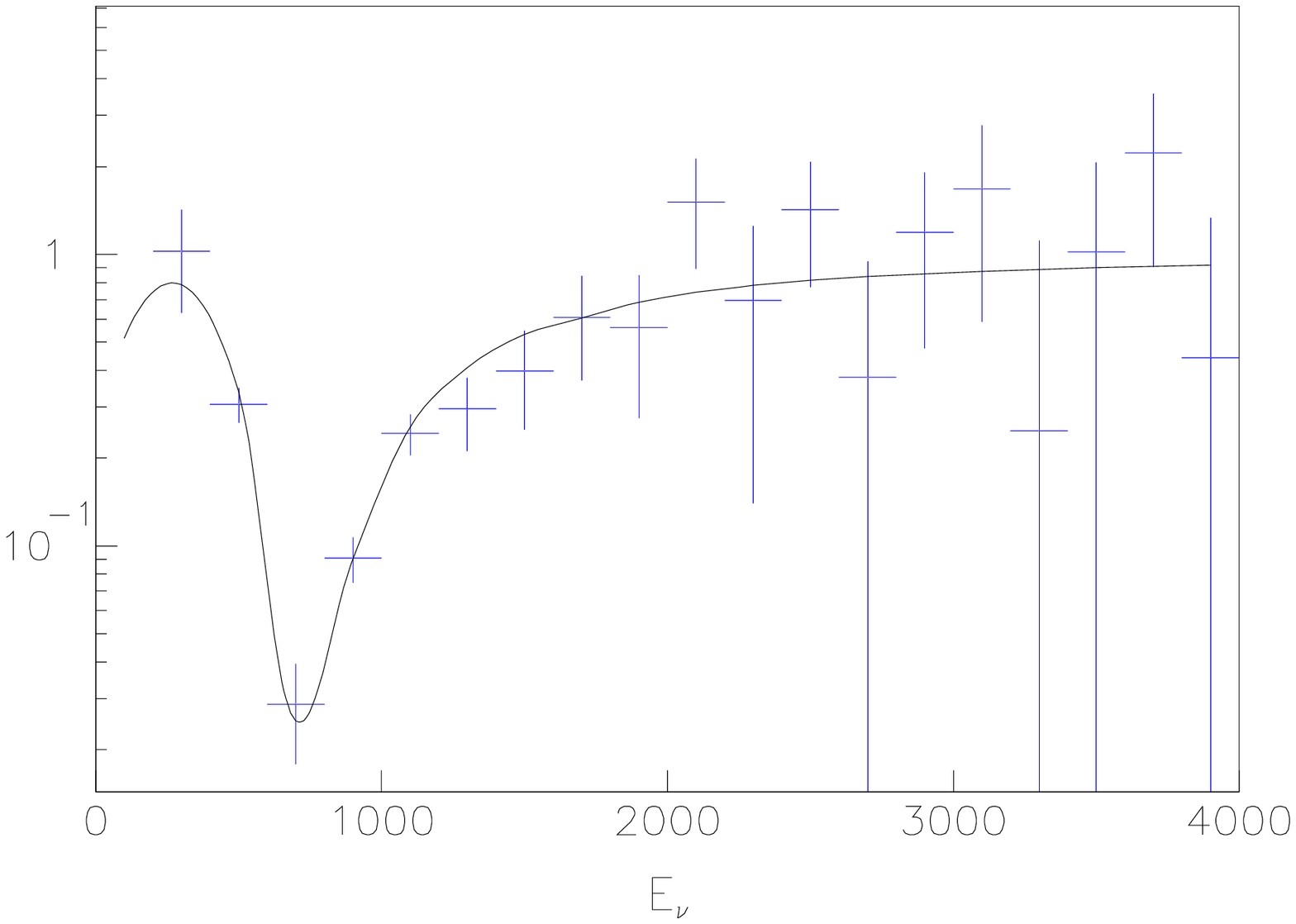,width=12cm}}
\vspace{-4mm}
\caption{\protect\footnotesize
The ratio of the measured spectrum with neutrino 
oscillation to the expected one without neutrino 
oscillation after subtracting the contribution of non QE-events. 
The fit result of the oscillation is overlaid.}
\label{numu:ratio}
\end{figure}

In the oscillation analysis, the true neutrino energy 
spectrum is extracted by subtracting the contribution 
of non-QE background events. 
To measure the oscillation parameters,
full Super-Kamiokande Monte Carlo events are generated and
the ratio between the ``measured'' spectrum at SK and the 
the expected one without oscillation is fitted by the function
of $P(\nu_\mu \rightarrow \nu_\mu)$ in Equation~\ref{eqn:Pmm} 
after subtracting the non-QE contribution.
Since the spectrum of non-QE events depends on the oscillation 
parameters, the non-QE spectrum are updated by the fit result
at each iteration of the fitting.
The survival probability of $P(\nu_\mu \rightarrow \nu_\mu)$ 
is shown in Figure~\ref{numu:ratio}, which gives the fit result of
$(\Delta m_{23}, \sin^22\theta_{23}) = \rm 
((2.96 \pm 0.04) \times 10^{-3} eV^2, 1.0 \pm 0.01)$. 
The oscillation pattern is clearly seen and the
$\sin^22\theta_{23}$ precision of 1~\%  
and the $\Delta m_{23}$ precision of $4\times 10^{-5}$~$\rm eV^2$
are expected.

\begin{figure}[!tb]
\renewcommand{\baselinestretch}{1}
\centerline{\epsfig{figure=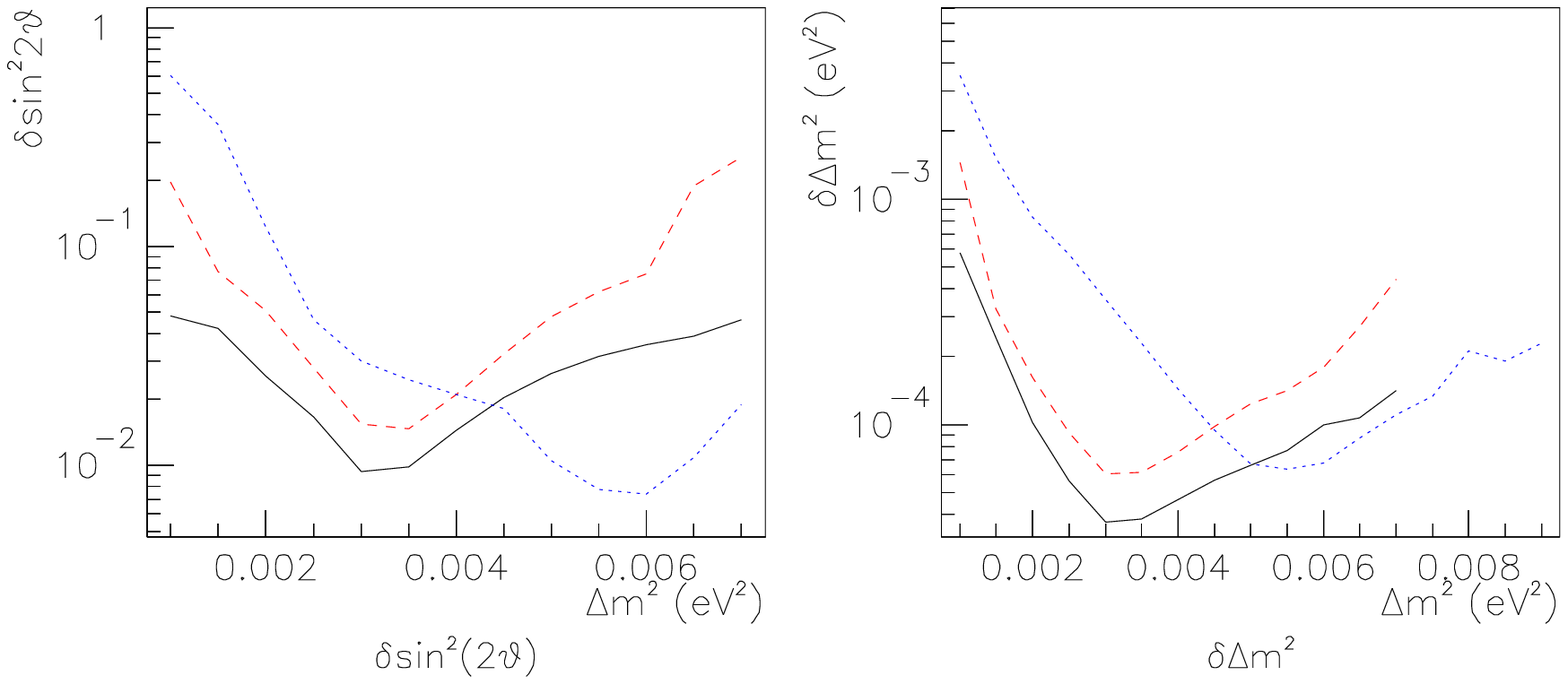,width=17cm}} 
\vspace{-7mm}
\caption{\protect\footnotesize
The final sensitivity of the neutrino oscillation parameters:
$\sin^22 \theta_{23}$ (left) and $\Delta m_{23}$ (right), as a function 
of true $\Delta m_{23}$ ($\rm eV^2$). The $\sin^2 2 \theta_{23}$ is 
set to 1.00. The result with OA2$^\circ$ is shown by the black (solid) line, 
LE$1.5 \pi$ by the red (dashed) line and LE$3 \pi$ by the blue
(dotted) line.}
\label{numu:result}
\end{figure}

Several beam configurations are studied in the range of 
$\Delta m_{23}$ between $1 \times 10^{-3}$ and 
$1 \times 10^{-2}$~$\rm eV^2$. 
The result is summarized in Figure~\ref{numu:result}.
With OA2$^\circ$, the maximum sensitivity to the oscillation parameters 
is achieved at $\Delta m_{23}=(3 \sim 3.5) \times 10^{-3}$~$\rm eV^2$. 

\begin{figure}[!tb]
\renewcommand{\baselinestretch}{1}
\centerline{\epsfig{figure=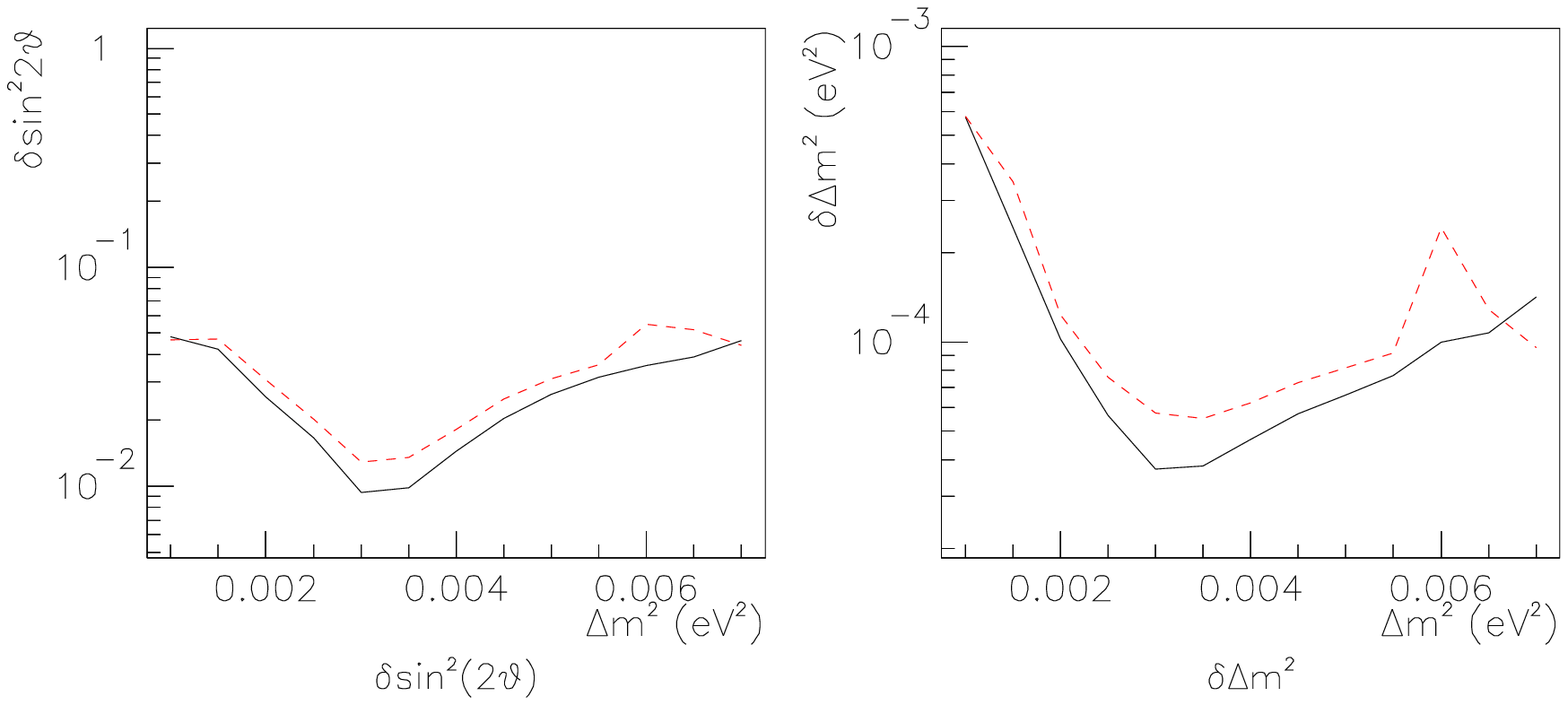,width=17cm}} 
\vspace{-7mm}
\caption{\protect\footnotesize
The final sensitivity of the neutrino oscillation parameters
in the case of $\sin^22 \theta_{23}=0.9$ shown by red (dashed) line:
$\sin^22 \theta_{23}$ (left) and $\Delta m_{23}$ (right), as a function 
of true $\Delta m_{23}$ ($\rm eV^2$). As a reference, 
the result of $\sin^22 \theta_{23} = 1.00$ is shown in black line (solid).}
\label{numu:result09}
\end{figure}

So far, we assumed $\sin^22\theta_{23}=1.0$.
In the case of $\sin^22\theta_{23} = 0.9$, which is the lower 
bound suggested by atmospheric neutrino result of Super-Kamiokande,
is shown in Figure~\ref{numu:result09}. 
The precision is slightly worse due to non-oscillated 
neutrino events at the oscillation maximum.

By selecting the bin at the oscillation maximum,
the disappearance signal dip is enhanced and thus
the contribution of the systematic uncertainties are largely
suppressed.
For example, the depth of the dip in Figure~\ref{numu:ratio},
which corresponds to $1-\sin^22\theta_{23}$,
is as small as 3\%.
Thus, a systematic uncertainty of 10\% in the flux normalization
(far/near ratio) contributes to $3\%\times 0.1=$0.3\% in
the $\sin^22\theta_{23}$ measurement.
Assuming 10\% systematic uncertainty in the far/near ratio,
which is similar to K2K's number of 6\%,
4\% uncertainty in the energy scale,
and 20\% uncertainty in the non-QE background subtraction,
the total systematic error is estimated to be less than 1~\% for 
$\sin^22\theta_{23}$ and less than $1\times 10^{-4}$~$\rm eV^2$ for 
$\Delta m_{23}$. 
The systematic uncertainties are expected to be reduced further
below the statistical uncertainties 
by the neutrino flux measurement using QE events and detailed non-QE
background measurement by the front detector,
by the pion production measurement,
and possibly a second near detector at a few km point which makes
the systematics in the far/near ratio negligible.

The overall sensitivity of the first phase is expected to be 1\%
in precision for $\sin^22\theta_{23}$ and
better than $1\times 10^{-4}$~$\rm eV^2$ 
for $\Delta m_{23}$.

\subsection{$\nu_e$ appearance search}
\hspace*{\parindent}
The JHF neutrino beam has small $\nu_e$ contamination
(0.2\% at the peak energy of OAB and NBB)
and the $\nu_e$ appearance signal is enhanced
by tuning the neutrino energy at its expected
oscillation maximum.
Reconstruction of the neutrino energy
provides oscillation pattern and
makes a positive identification of the $\nu_e$
appearance signal.
Thus, JHF-Kamioka neutrino experiment has an excellent opportunity 
to discover $\nu_e$ appearance and thus measure $\theta_{13}$.
In this section, the sensitivity on $\nu_e$ appearance
is described based on the full Monte Carlo simulations
and analysis of Super-Kamiokande and K2K experiments.

\subsubsection{Background sources and event selection criteria}
\hspace*{\parindent}
The process of the $\nu_e$ appearance signal 
is searched for in the QE interaction. 
Since the proton momentum from the QE interaction is usually below 
the \v{C}erenkov threshold, the signal has only  
a single electro-magnetic shower (single ring e-like).
The possible background processes are 
$\nu_\mu\rightarrow \mu$
with e/$\mu$ misidentification, $\nu_e$ contamination,
and $\pi^0$ background.
The background from $\mu$ misidentification is found to be 
negligible due to excellent e/$\mu$ separation of Super-Kamiokande.
The $\nu_e$ contamination is as small as 0.2-0.3\%.
A $\pi^0$ produced by neutral current
and inelastic charged current processes is a possible background 
when one of the photon is missed or 2 photons overlaps.

The standard Super-Kamiokande atmospheric neutrino analysis
requirements are used to select a single ring e-like event:
single ring, electron like (showering), visible energy greater than 100~MeV, 
and no decay electrons.
The electron identification eliminates all of the
$\mu$ background, and the decay electron cut further eliminates
inelastic charged current processes associated with $\pi^0$.
Reduction of number of events by the
``standard'' 1ring e-like cut for charged and neutral current
events are listed in
\tablename~\ref{nueapp:cuteff_oa2} 2).
The remaining background events at this stage 
are predominantly from single
$\pi^0$ production through neutral current interaction and
a $\nu_e$ contamination.
\begin{table}
\renewcommand{\baselinestretch}{1}
\begin{center}
\caption{\protect\footnotesize Number of events and reduction efficiency
of ``standard'' 1ring e-like cut and $\pi^0$ cut for 5 year exposure
($5 \times 10^{21}$ p.o.t.) OA$2^\circ$.
For the calculation of oscillated $\nu_e$,$\Delta m^2=3\times10^{-3}$~eV$^2$
and $\sin^22\theta_{\mu e}=0.05$ is assumed. \label{nueapp:cuteff_oa2}}
\begin{tabular}{lrrrr}
\hline
OAB $2^\circ$ & $\nu_\mu$ C.C. & $\nu_\mu$ N.C. & Beam $\nu_e$ & Oscillated $\nu_e$ \\ 
\hline 
1) Generated in F.V. & 10713.6 & 4080.3 & 292.1 & 301.6 \\
2) 1R e-like & 14.3 & 247.1 & 68.4 & 203.7 \\
3) e/$\pi^0$ separation & 3.5 & 23.0 & 21.9 & 152.2 \\
4) 0.4~GeV$<E_{rec}<1.2$~GeV & 1.8 & 9.3 & 11.1 & 123.2  \\
\hline
\end{tabular}
\end{center}
\end{table}

For neutrino energies below 1~GeV,
the energy of produced $\pi^0$ is low and
thus the probability of the two photons to merge to one
is small. 
The limitation comes from asymmetric decay of $\pi^0$
with one high and one low photon energies,
where the lower energy photon tends to be hidden
under the scattered light\footnote{About 10-20\% of the
light are scattered in the water before reaching the 
photomultiplier tubes causing broadly distributed 
background light.}
of the higher energy photon.
In order to recover this hidden lower energy photon
and further suppress the $\pi^0$ background,
the photomultiplier hit pattern
including scattered light is fitted
under the hypotheses of single and that of double electro-magnetic rings.
\figurename~\ref{nueapp:qdist} shows 
distributions of four characteristic quantities
that separates signal $\nu_e$ events from $\pi^0$
background events as follows:
\begin{figure}
\renewcommand{\baselinestretch}{1}
\centerline{
\epsfig{file=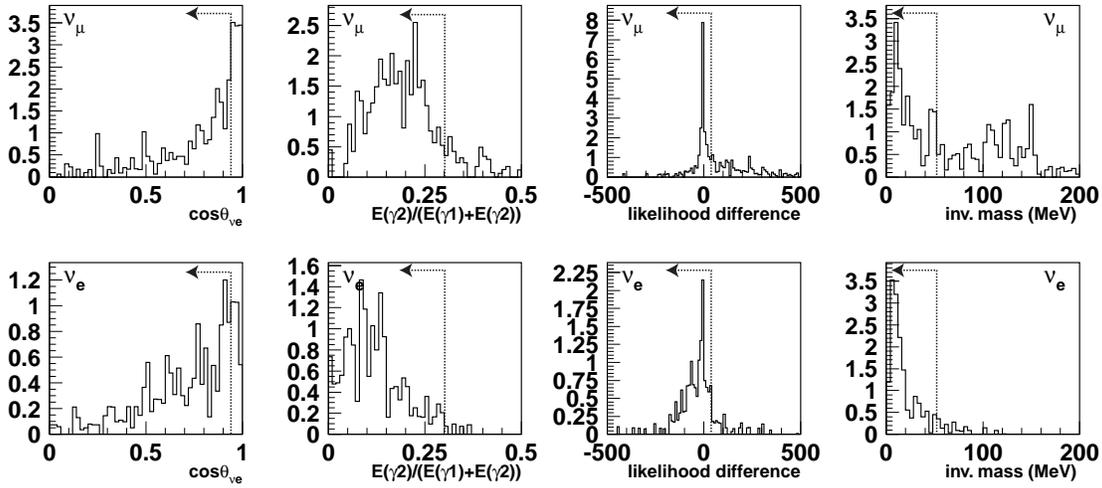,width=16cm}
}
\vspace{-8mm}
\caption{\protect\footnotesize
Distributions of the four quantities,
used in the $e/\pi^0$ separation.
The beam is the wide band beam and events are
after the single-ring e-like selection.
Upper histograms corresponds to $\nu_\mu$ background events 
and the lower histograms correspond to the $\nu_e$ signal events.
The arrows in the figure show the cut positions used in the analysis.}
\label{nueapp:qdist}
\end{figure}
\begin{itemize}
\item Angle between $\nu$ and e ($\cos\theta_{\nu e}$):\\
Some fraction of $\pi^0$ background has a steep forward peak,
which is likely due to coherent $\pi^0$ production.
Those events in the extreme forward direction are rejected.
\item Energy fraction of lower energy ring 
($\frac{E(\gamma_2)}{E(\gamma_1)+E(\gamma_2)}$)\\
The $\nu_e$ signal tends to have a low energy second ring
which is either a fake ring or a ring due to bremsstrahlung.
Those events with the large energy fraction are rejected.
\item Difference between double and single ring likelihood:\\
Single ring like events are selected. 
\item Invariant mass of 2 photons:\\
The $\pi^0$ background shows a peak at 135~MeV whereas
the $\nu_e$ signal shows small invariant mass.
Those events with large invariant mass are rejected.
\end{itemize}
\tablename~\ref{nueapp:cuteff_oa2} 3) lists the
number of events after this $e/\pi^0$ separation.
An order of magnitude extra rejection (247.1/23) in the $\nu_\mu$ neutral
current background is achieved with 152.2/203.7=75\% in signal acceptance.

\subsubsection{Oscillation sensitivity}
\begin{figure}
\renewcommand{\baselinestretch}{1}
\centerline{
\epsfig{file=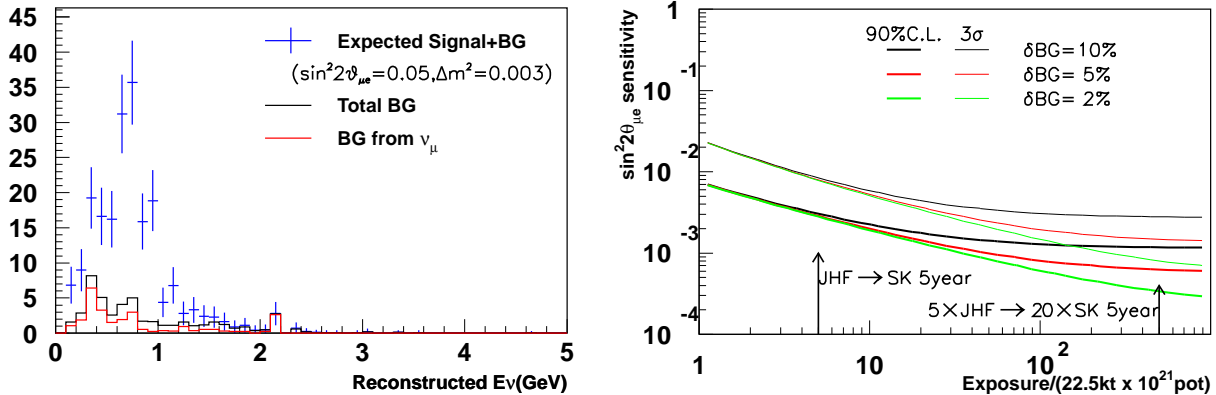,width=16cm}
}
\vspace{-4mm}
\caption{\protect\footnotesize
Left:Expected reconstructed neutrino energy distributions of
expected signal+BG, total BG, and BG from $\nu_\mu$ interactions
for 5 years exposure of OA2$^\circ$.
Right: Expected (thick lines:) 90\%CL sensitivity and (thin lines:)
3$\sigma$ discovery contours as the functions of exposure time of 
OA2$^\circ$. In left figure, expected
oscillation signals are calculated with the oscillation parameters:
$\Delta m^2 = 3\times10^{-3}$~eV$^2$,$\sin^22\theta_{\mu e}=0.05$.
In right figures, Three different contours correspond to 
10\%, 5\%, and 2\% uncertainty in the background estimation.}
\label{nueapp:sens_oa2.0}
\end{figure}

\hspace*{\parindent}
\figurename~\ref{nueapp:sens_oa2.0} (left)
shows the reconstructed neutrino energy distributions 
for 5~years exposure of JHF OA2$^\circ$.
The oscillation parameters of
$\Delta m^2 = 3\times10^{-3}$~eV$^2$ 
and $\sin^22\theta_{\mu e}=0.05$ are assumed.
A clear appearance peak is seen at the oscillation maximum
of $E_\nu \sim$0.75~GeV.
By integrating the number of events in the energy range between
0.4 and 1.2~GeV, 90\% and 3$\sigma$ limits are derived
as a function of the exposure time
(the right plot of \figurename~\ref{nueapp:sens_oa2.0}).
The reach of the first phase is as good as 
$\sin^22\theta_{\mu e}=0.003$ at 90\% confidence level.
The systematic uncertainty in background subtraction
is chosen to be 2\%, 5\% and 10\%.
Though the systematic uncertainty is not important
in the first phase, it becomes significant in the second 
phase.

\begin{figure}[!tb]
\renewcommand{\baselinestretch}{1}
\centerline{
\epsfig{file=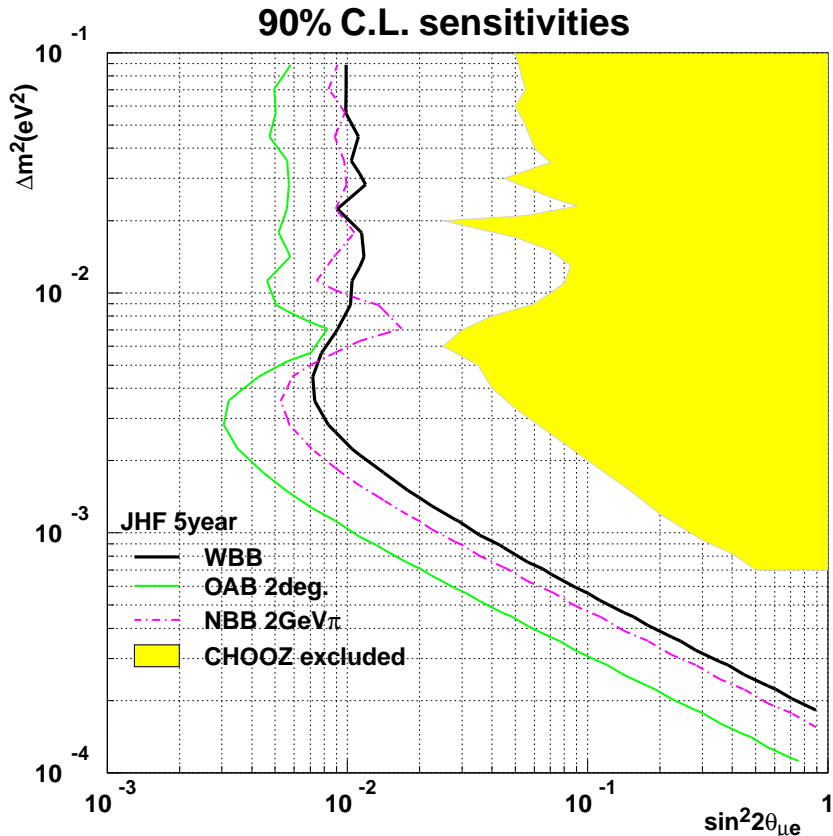,width=12cm}
}
\vspace{-7mm}
\caption{\protect\footnotesize
The 90\% C.L. sensitivity contours for 5 years exposure of
WBB, OA2$^\circ$, and LE$2 \pi$ configurations. 
The 90\% C.L. excluded region of CHOOZ is plotted as
a comparison. For CHOOZ contour, maximum mixing of $\sin^2\theta_{23}=0.5$
is assumed to convert from $\sin^22\theta_{13}$ to $\sin^22\theta_{\mu e}$.}
\label{nueapp:contours}
\end{figure}

\figurename~\ref{nueapp:contours} shows 90\%C.L. contours for
5 year exposure of each beam configuration assuming 
10\% systematic uncertainty in background subtraction.

The best sensitivity at around $\Delta m^2=3\times10^{-3}$~eV$^2$
is given by OA2$^\circ$ and the sensitivity is
$\sin^22\theta_{\mu e}=0.003$ at 90\%C.L. 
If the $\Delta m^2$ is significantly larger or smaller than 
$3\times10^{-3}$~eV$^2$, the sensitivity can be optimized by
adjusting the neutrino energy as described in section~\ref{beam}.
In the predicted range of 
$1.6\times10^{-3}$~eV$^2<\Delta m^2<4\times10^{-3}$~eV$^2$
by the Super-Kamiokande,
the  sensitivity is 
better than $\sin^22\theta_{\mu e}=0.005$ 
or $\sin^22\theta_{13}=0.01$ at 90\% confidence level.

\subsection{Confirmation of $\nu_{\mu}\rightarrow\nu_{\tau}$ 
in atmospheric neutrino observation}
\hspace*{\parindent}
Neutral current (NC) interaction detects the sum of 
$\nu_{\mu}\rightarrow\nu_e,\nu_{\mu}$, and $\nu_{\tau}$
oscillations\footnote{According to the LEP result,
there are only 3 light active neutrinos.}.
Therefore, NC measurement combined with 
$\nu_{\mu}\rightarrow\nu_e$ and $\nu_{\mu}\rightarrow\nu_{\mu}$
measurements provide indirect measurement of the
$\nu_{\mu}\rightarrow\nu_{\tau}$ oscillation.
The NC measurement also provides a constraint 
on the existence of a sterile neutrino, $\nu_s$, since 
the $\nu_{\mu}\rightarrow\nu_s$ oscillation
results in reduction of NC interactions.

In the JHF neutrino beam energy of $\sim$1~GeV, 
the dominant $\nu_\mu$ NC interactions are single $\pi$ productions.
Among those, single $\pi^0$ production process has a unique
signature that only electromagnetic showers
made by decay photons are observed without any other
activities. Therefore, we focus on the single $\pi^0$
production mode.

A background from charged current (CC) interactions with a $\pi^0$
are separated by identifying at least one muon in the final state.
Muons are identified by detecting an extra \v{C}erenkov ring or the Cherenkov photons,
or by detecting the decay electrons.
In the Super-Kamiokande analysis, the efficiency of detecting
the decay electrons is better than 80\%.
Another background is from NC interactions by high energy
neutrinos which are out of the oscillation maximum.
Though we cannot reconstruct the initial neutrino energy
for the NC interaction, the total energy deposit 
in the detector, called ``visible energy'', is used to reject  
high energy events.
When the high energy neutrino produces multiple pions,
which is the dominant process at high energy,
they are further suppressed by vetoing on the extra pion
rings  as well as decay electrons from the 
${\pi^+\rightarrow \mu^+ \nu_\mu}, {\mu^+ \rightarrow e^+ \bar{\nu_\mu} \nu_e}$
decay chain.

Thus, the selection criteria are defined as follows:
1) the event must be fully contained,
2) visible energy (electron equivalent energy) must be greater
   than 100~MeV and less than 1500~MeV,
3) the number of rings in an event must be less than three, 
4) all the rings must be electron-like and
5) no decay electron is identified.

After applying these selection criteria, 
about 315 events are expected to be observed in one year
without neutrino oscillation for WBB.
Out of 315 events, 88\% are from NC interactions, 
9\% are from $\nu_\mu$ CC interactions, and 3\% are from $\nu_e$ CC interaction 
intrinsically in the beam.
The remaining 9\% events from $\nu_\mu$ CC interactions is hard
to be eliminated because the muon is captured by oxygen and 
does not decay. The results are summarized in
Table~\ref{nc:tab:sum} together with the other beam configurations.
\begin{table}
\renewcommand{\baselinestretch}{1}
\begin{center}
\caption{\protect\footnotesize
Summary of the event rate of the NC candidates.
$f(X)$ is the fraction of the events from $X$ interaction.} \label{nc:tab:sum}
\begin{tabular}{lccccc}
\hline
Beam & \#NC Events & Beam exposure& $f(\nu_\mu \rm NC)$ 
& $f(\nu_\mu \rm CC)$ &   $(\nu_e \rm CC)$ \\ 
\hline 
WBB         & 315 & 1 years & 0.88 & 0.09 & 0.03 \\
LE2$\pi$    & 250 & 5 years & 0.80 & 0.13 & 0.07 \\
OA2$^\circ$ & 700 & 5 years & 0.84 & 0.09 & 0.07 \\
\hline
\end{tabular}
\end{center}
\end{table}

\begin{figure}
\renewcommand{\baselinestretch}{1}
\centerline{(a)\hspace*{0.3\textwidth}(b)\hspace*{0.3\textwidth}(c)}
\centerline{
\epsfig{file=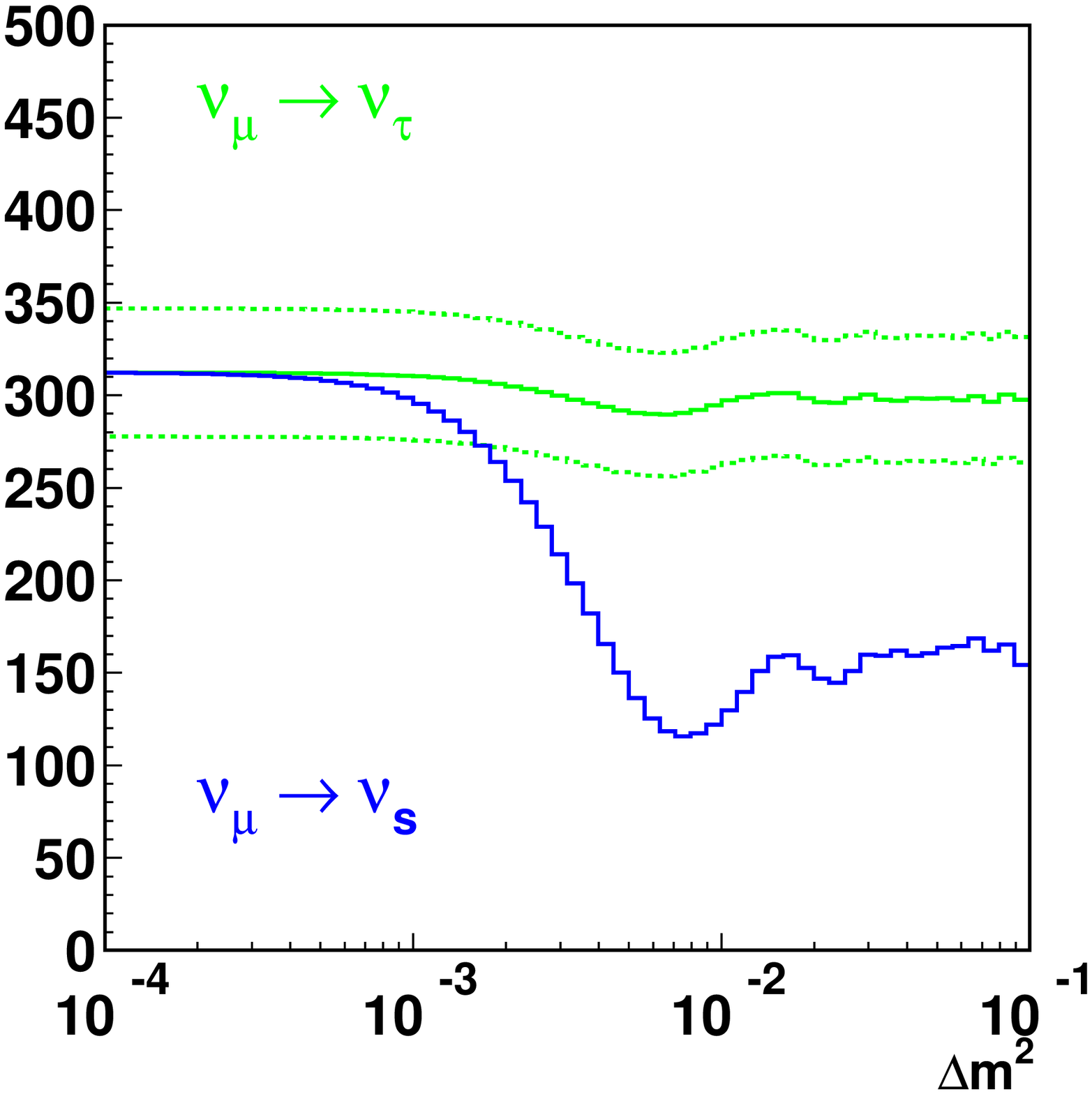,width=0.3\textwidth}
\epsfig{file=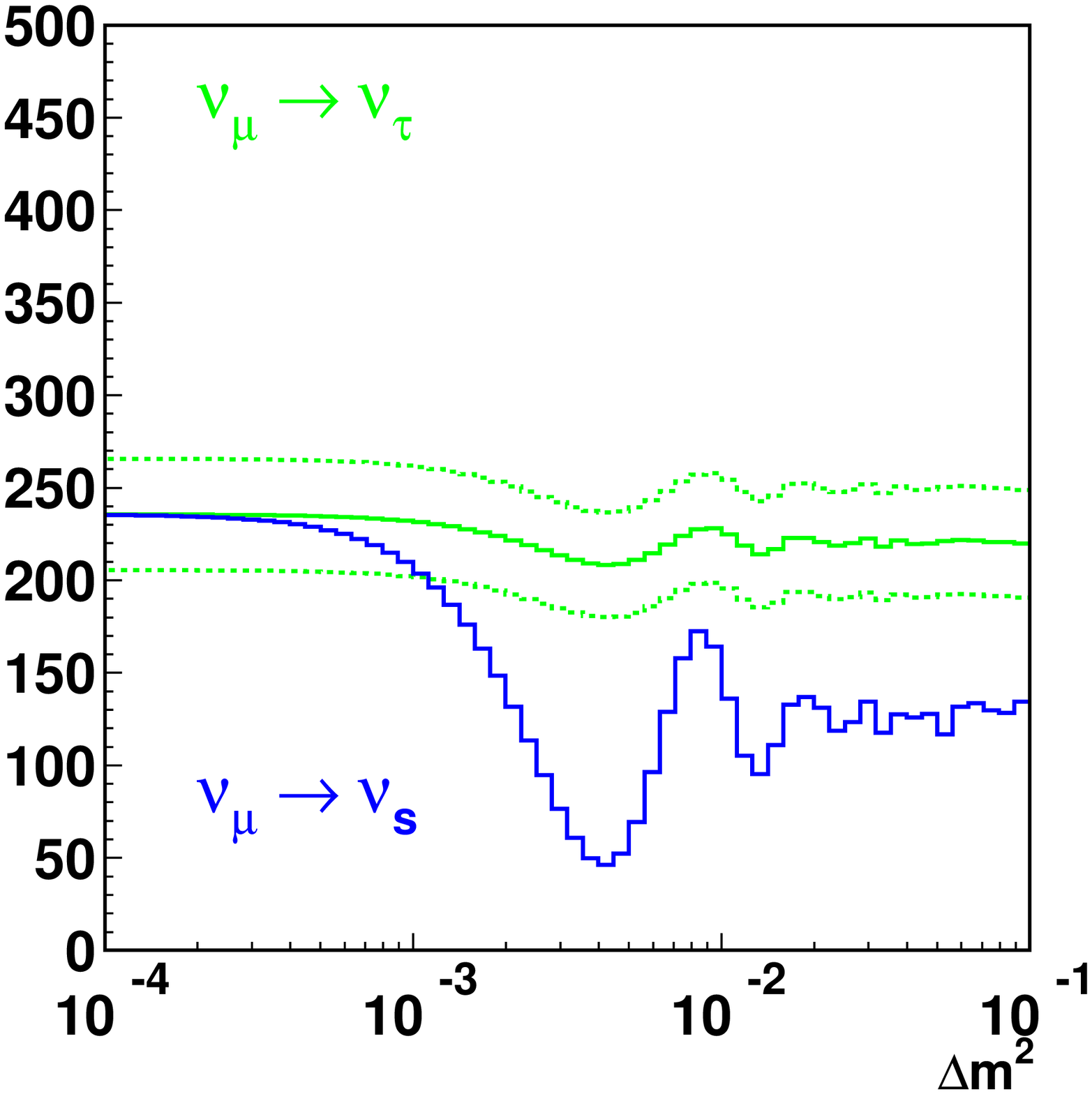,width=0.3\textwidth}
\epsfig{file=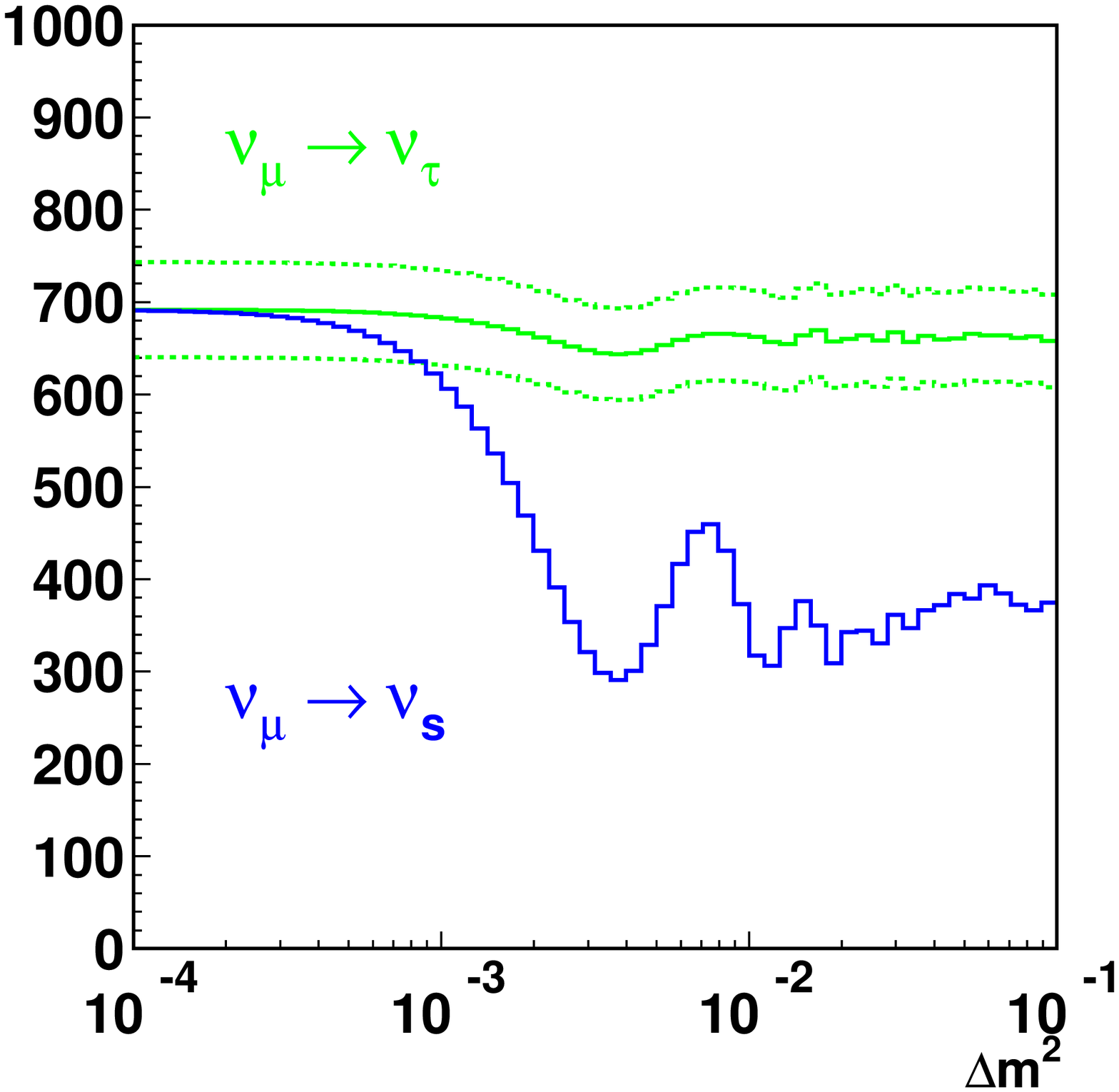,width=0.3\textwidth}
}
\vspace{-4mm}
\caption{\protect\footnotesize
Expected number of events with various $\Delta m^2$
for (a) 1 year of WBB,
(b) 5 years of LE2$\pi$
and (c) 5 years of OA2$^\circ$.
The solid lines show the expected numbers of events
assuming $\nu_\mu \rightarrow \nu_\tau$
or $\nu_\mu \rightarrow \nu_s$. The dotted lines show the
90\% C.L. regions of $\nu_\mu \rightarrow \nu_\tau$ oscillation.}
\label{TAUSTERILE}
\end{figure}

The expected numbers of events as a function of $\Delta m^2$ are 
shown in Figure~\ref{TAUSTERILE}.
In the figures, maximal oscillations, $\sin^22\theta_{23}$=1.0, is assumed.
Although full mixing is assumed, expected number of events
does not become 0 even at the deepest dip point. This is due to
the NC interactions of high energy neutrinos.
The dotted lines in each figure correspond to the 90\% C.L.
limit for $\nu_\mu \rightarrow \nu_\tau$ oscillations
assuming a systematic uncertainty of 5\%. 
The expected numbers of events for
$\nu_\mu \rightarrow \nu_\tau$ and for $\nu_\mu \rightarrow \nu_s$
are clearly separated if the $\Delta m^2$ is larger than 
$2 \times 10^{-3}$ for WBB,
$1.5 \times 10^{-3}$ for LE2$\pi$ and $1\times 10^{-3}$ for OA2$^\circ$.

\section{Physics in the future extension with Hyper-Kamiokande}
\hspace*{\parindent}
In the 2nd phase of the JHF-Kamioka neutrino experiment,
the proton intensity is planned to go up to 4~MW~\cite{mori}.
The pion (or neutrino) production target will also be
upgraded to a liquid metal target to accept the 4~MW beam.
The shielding of the decay pipe will be designed to
accomodate such a beam.

\begin{figure}[!tb]
\renewcommand{\baselinestretch}{1}
\centerline{\epsfig{file=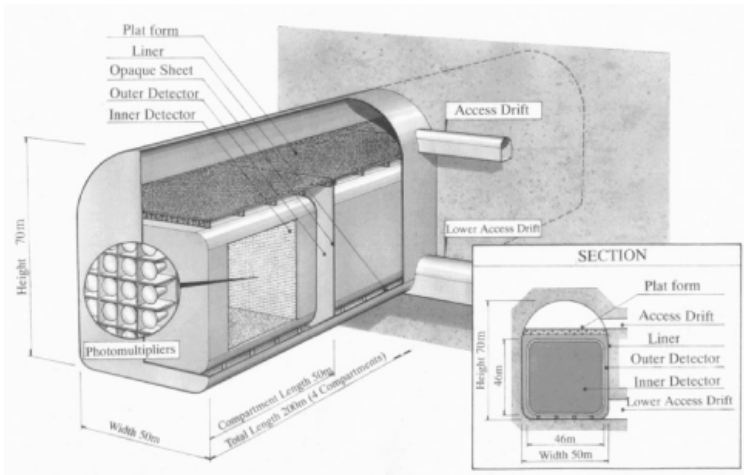,width=10cm}}
\vspace{-4mm}
\caption{\protect\footnotesize
Schematic view of the \Hyper-K{} detector.}
\label{fig:Hyper-K}
\end{figure}

As for the far detector,
\Hyper-K{} detector is proposed as a next generation large water 
\v{C}erenkov detector \cite{hyperk} at Mozumi zinc mine in Kamioka,
where the \Super-K{} detector is located.
Schematic view of one candidate detector design is shown
in Figure \ref{fig:Hyper-K}.
A large water tank is made from several 
50 m \(\times\) 50 m \(\times\) 50 m sub-detectors.  
The tank will be filled with pure water and 
photomultiplier tubes (PMTs) are instrumented on all
surfaces of sub-detectors.
The fiducial volume of each sub-detector is about 70~kt and 1~Mt
volume is achieved by 14 sub-detectors.  
The 2.0 m thick outer detectors
completely surround the inner sub-detectors and  
the outer region is also instrumented with PMTs.
The primary function of the outer detectors is to
veto cosmic ray muons and to help identify contained events.
The Kamioka site satisfies the conditions required for constructing
large water \v{C}erenkov detectors: easy access to underground,
clean water, hard and uniform rock, and infrastructure/technology
for excavation. The overburden of the \Hyper-K{}
is expected to be somewhere between 1900 and 2700 meter-water-equivalent,
depending on the actual location of the detector.

With these upgrades in both accelerator ($\times 5$)
and detector ($\times 40$),
the statistics is expected to increase by
a factor of 200. The goal of the second phase is
\begin{itemize}
\vspace{-1mm}
   \item sin$^2 2\theta_{13}$ sensitivity below $10^{-3}$   
\vspace{-1mm}
   \item CP phase $\delta$ measurement down to 10-20 degrees
\vspace{-1mm}
   \item Test of the unitarity triangle in the lepton sector
\vspace{-1mm}
   \item Search for Proton decay: $p\rightarrow K^+\bar\nu, e^+\pi^0$
\end{itemize}
\vspace{-1mm}

\subsection{Discovery potential of CP violation in the lepton sector}
\hspace*{\parindent}
If $\nu_\mu \rightarrow \nu_e$ is not observed in the first phase,
another order of magnitude improvement in $\sin^22\theta_{\mu e}$ 
sensitivity to better than $10^{-3}$ will be performed in the second phase 
(Fig~\ref{nueapp:sens_oa2.0}).
Systematic uncertainty in background subtraction becomes important
in the 2nd phase. 
Enhancement of the signal at the oscillation maximum 
and capability of measuring the background
by the side-band of the oscillation pattern in the reconstructed
neutrino energy distribution
provide an excellent handle in controlling the systematic uncertainty.

If the large mixing angle solution of the solar neutrino
deficit, which is the current favored solution, and
if $\nu_\mu \rightarrow \nu_e$ is observed in the first phase,
there is a good chance of observing CP violation
in the 2nd phase.
The CP asymmetry is calculated  as
\begin{eqnarray}
A_{CP}&=&
{P(\nu_\mu \rightarrow \nu_e)
  -P(\bar\nu_\mu \rightarrow \bar\nu_e) \over
 P(\nu_\mu \rightarrow \nu_e) 
  +P(\bar\nu_\mu \rightarrow \bar\nu_e)} =
{\Delta m_{12}^2 L \over 4E_\nu}
\cdot {\sin2\theta_{12} \over \sin\theta_{13}} \cdot \sin\delta  \nonumber 
\end{eqnarray}
\begin{figure}[!tb]
\renewcommand{\baselinestretch}{1}
\vspace{-1cm}
\centerline{\epsfig{file=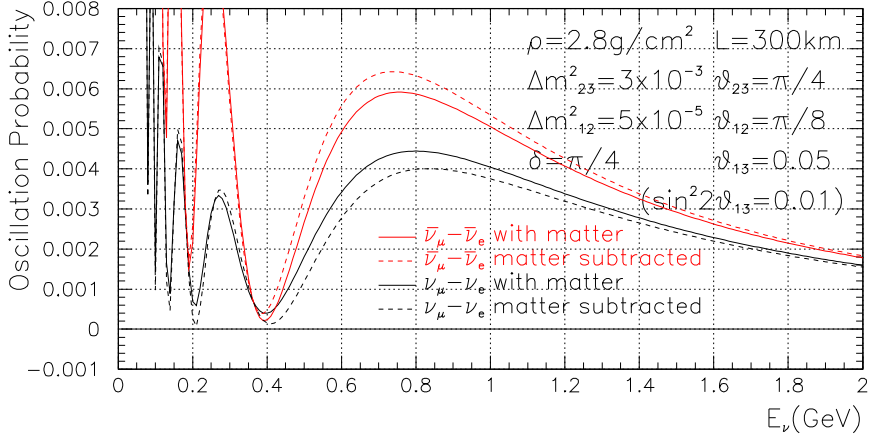,width=12.0cm}}
\vspace{-7mm}
\caption{\protect\footnotesize
Oscillation probabilities for
$\nu_\mu\rightarrow \nu_e$(black) and
$\bar\nu_\mu\rightarrow \bar\nu_e$(red).
The solid curves includes asymmetry due to matter effect.
For the dashed curves, the matter effect is subtracted
and the difference between $\nu_\mu\rightarrow \nu_e$(black) and
$\bar\nu_\mu\rightarrow \bar\nu_e$(red)
are all due to CP effect.
}
\label{fig:oscillate_cp1}
\end{figure}
By choosing low energy neutrino beam at the oscillation maximum
(E$\sim$0.75~GeV and L$\sim$295~km for JHF)
the CP asymmetry is enhanced as 1/E.
Taking the central value of LMA,
namely $\theta_{12}= \pi/8$ and $\Delta m^2_{12}= 5\times 10^{-5}$,
and $\sin^22\theta_{13}=0.01$ (1/10 of CHOOZ limit) and
$\delta = \pi/4$ (half of the maximum CP effect),
$A_{CP}$ becomes as large as 25\%.
Matter effect, which creates fake CP asymmetry, increases
linearly with the neutrino energy.
Because of the use of low energy neutrinos, the fake asymmetry
due to matter effect is small for the JHF-Kamioka experiment.
Figure~\ref{fig:oscillate_cp1} shows
oscillation probabilities for
$\nu_\mu\rightarrow \nu_e$(black) and
$\bar\nu_\mu\rightarrow \bar\nu_e$(red)
for the central LMA parameters as discussed above.
The solid curves includes asymmetry due to matter effect.
For the dashed curves, the matter effect is subtracted
and the difference between $\nu_\mu\rightarrow \nu_e$(black) and
$\bar\nu_\mu\rightarrow \bar\nu_e$(red)
are all due to CP effect.
The CP asymmetry is as large as 25\% and the fake asymmetry
due to matter effect
is only 5-10\% as discussed above.

\subsubsection{Sensitivity to CP violation and the unitarity triangle} \label{cp}
\hspace*{\parindent}
\begin{figure}[!tb]
\renewcommand{\baselinestretch}{1}
\centerline{\epsfig{file=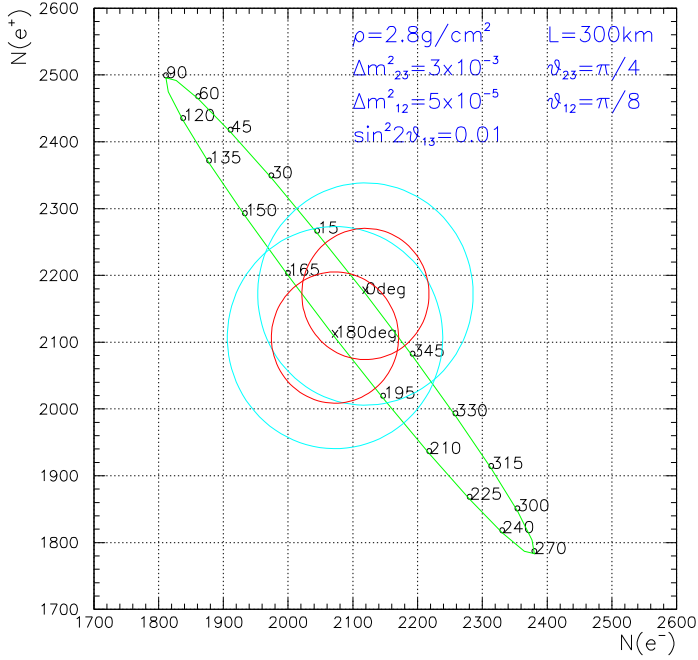,width=10cm}}
\vspace{-7mm}
\caption{\protect\footnotesize
Numbers of $\nu_e$ and $\bar\nu_e$ appearance events
including those from backgrounds.
Two circles indicate 3$\sigma$ contour (blue)
and 90\% confidence level (red) contours.}
\label{fig:cp_sens}
\end{figure}
\begin{figure}[!tb]
\renewcommand{\baselinestretch}{1}
\centerline{
\epsfig{file=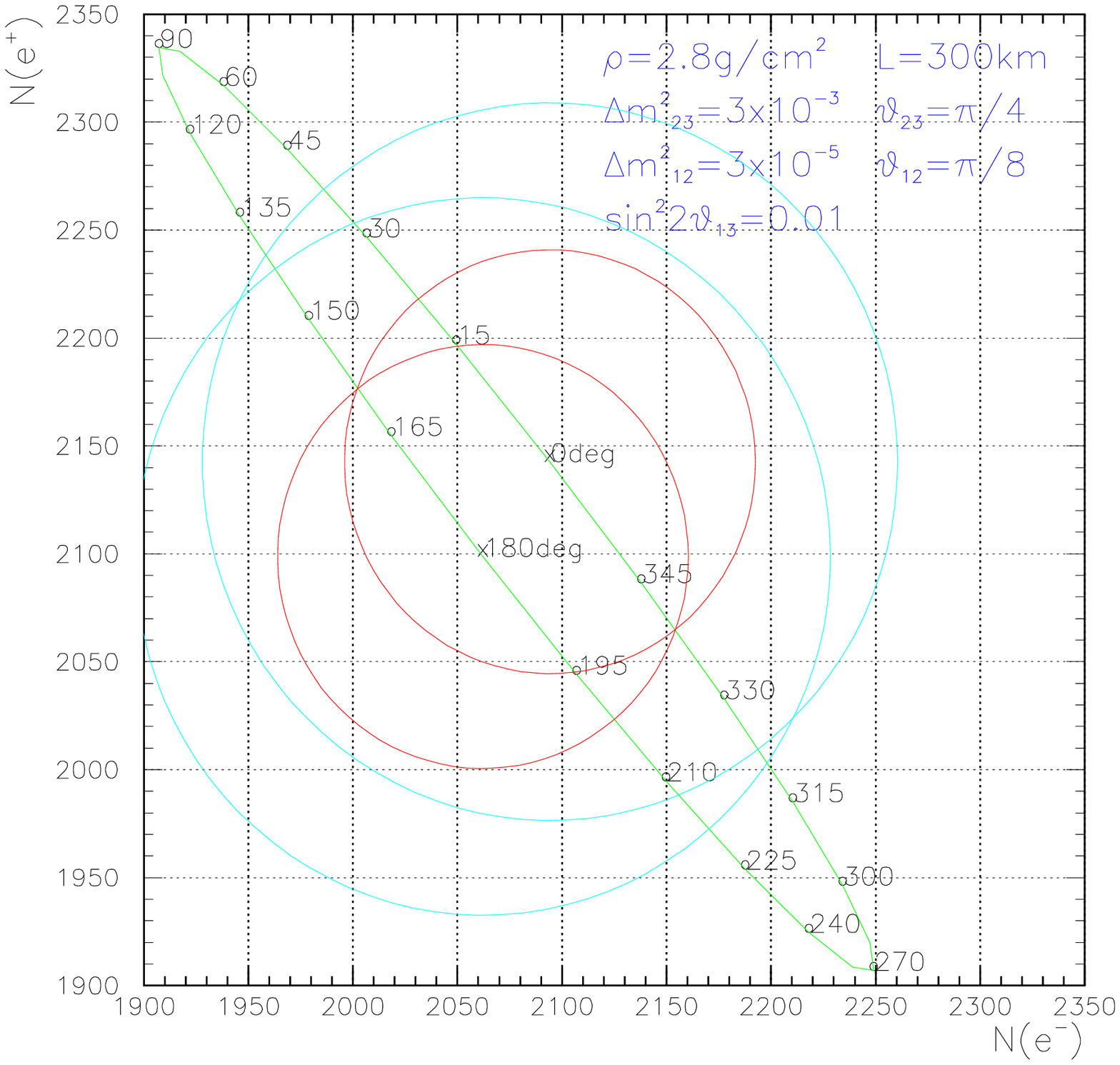,width=5cm}
\epsfig{file=fig/cp.3s.5.30.1.eps,width=5cm}
\epsfig{file=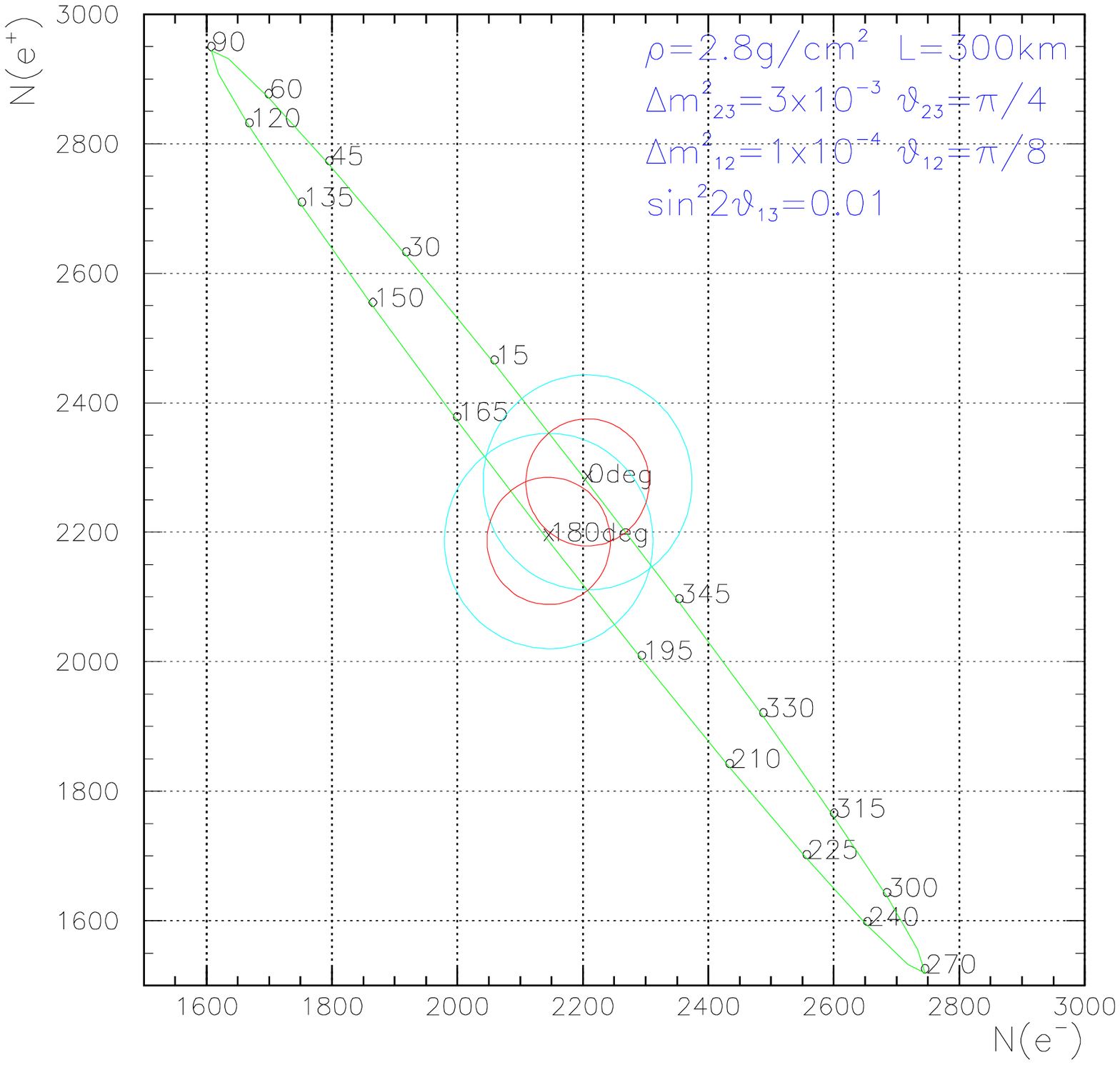,width=5cm}}
\vspace{-7mm}
\caption{\protect\footnotesize
CP sensitivity
for $\Delta m^2_{12}=3\times 10^{-5}$,  
$5\times 10^{-5}$ and $10\times 10^{-5}$.
}
\label{fig:cp_m12}
\end{figure}
\begin{figure}[!tb]
\renewcommand{\baselinestretch}{1}
\centerline{
\epsfig{file=fig/cp.3s.5.30.1.eps,width=5cm}
\epsfig{file=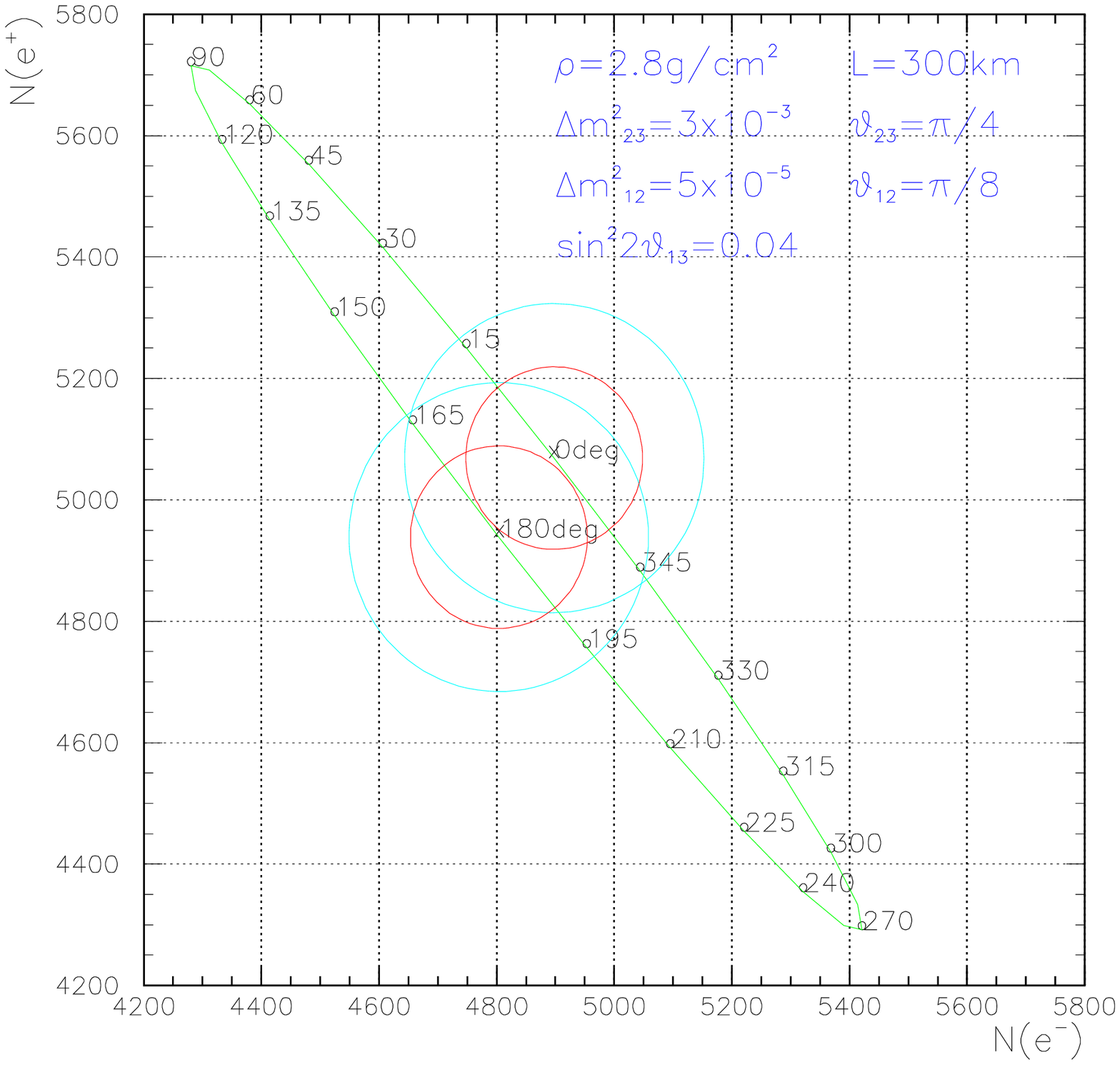,width=5cm}
\epsfig{file=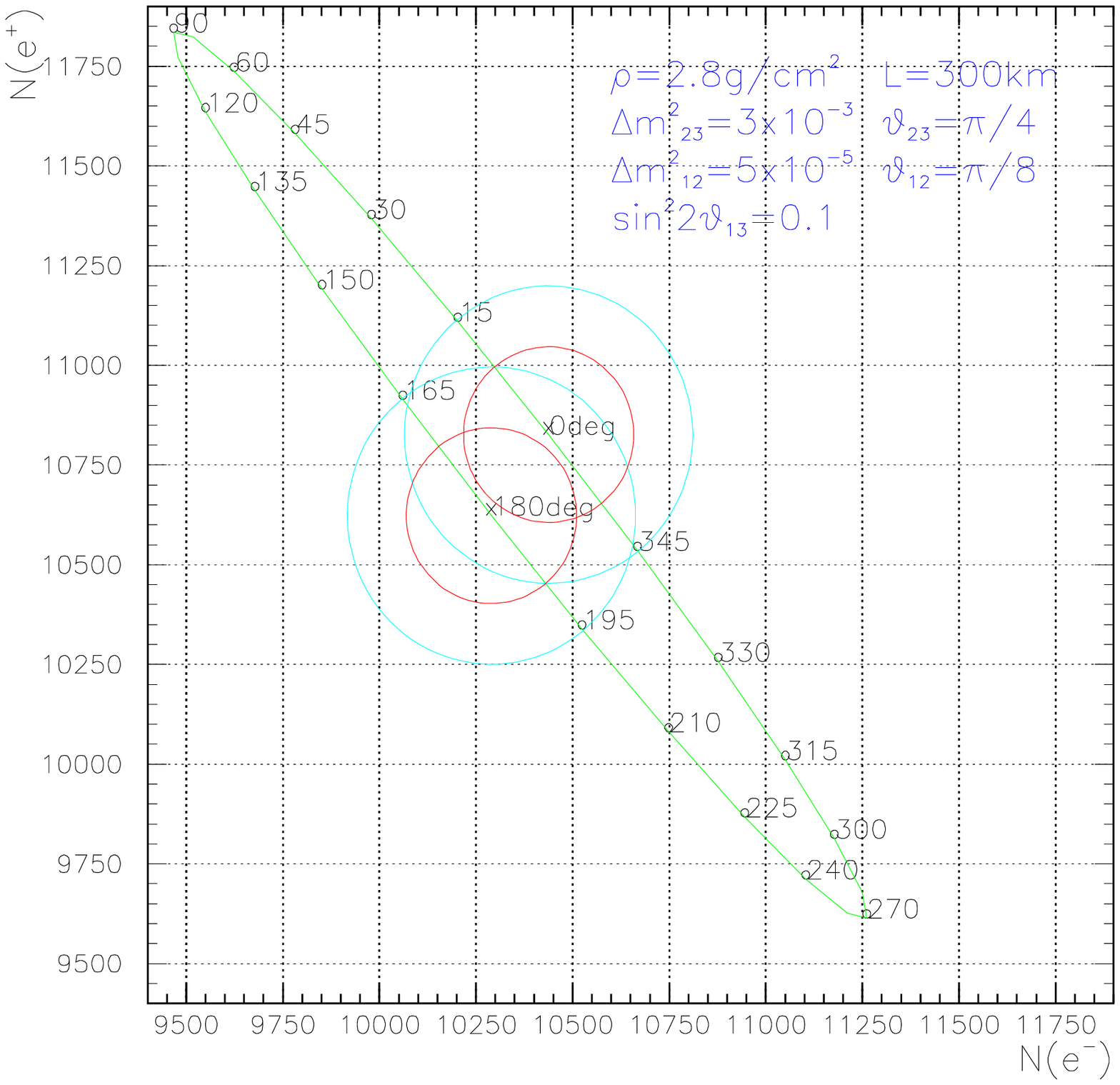,width=5cm}
}
\vspace{-7mm}
\caption{\protect\footnotesize
CP sensitivity
for $\sin^22\theta_{13}=$0.01, 0.04, and 0.10.  
}
\label{fig:cp_t13}
\end{figure}

Figure~\ref{fig:cp_sens} shows
the numbers of $\nu_e$ and $\bar\nu_e$ appearance events
including those from backgrounds after 6 years of $\bar\nu_\mu$
and 2 years of $\nu_\mu$ running in the 2nd phase of the
JHF-Kamioka neutrino experiment.
Numbers on the plots indicates CP phase $\delta$ in degrees.
CP phase at 0 degrees and 180 degrees correspond to 
no CP violation.
3 sigma discovery is possible for $|\delta|>20^\circ$. 

\begin{figure}[!tb]
\renewcommand{\baselinestretch}{1}
\vspace{-1cm}
\centerline{
\epsfig{file=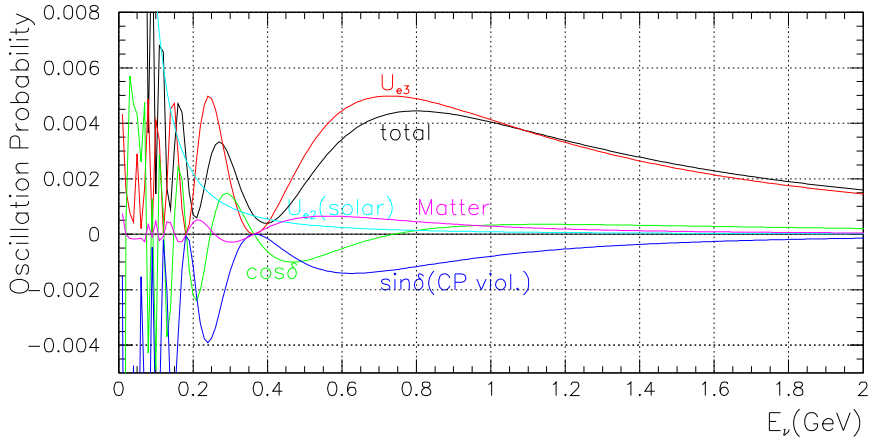,width=12cm}}
\vspace{-7mm}
\caption{\protect\footnotesize
Contribution of different component to
$\nu_\mu\rightarrow\nu_e$ appearance.}
\label{fig:unitary}
\end{figure}

Figure~\ref{fig:cp_m12} shows the CP sensitivity
for $\Delta m^2_{12}=3\times 10^{-5}$,  
$5\times 10^{-5}$ and $10\times 10^{-5}$.
Because  CP asymmetry ($A_{CP}$) increases
linearly with $\Delta m^2_{12}$,
CP sensitivity goes up linearly 
with $\Delta m^2_{12}$.
For small $\theta_{13}$,
CP asymmetry increases as $1/\sin\theta_{13}$,
but the statistics are down as $\sin^2\theta_{13}$.
As a result, CP sensitivity does not depend much
on the value of $\theta_{13}$ (for $\sin^22\theta_{13}>0.01$)
as shown in Figure~\ref{fig:cp_t13}.
Figure~\ref{fig:unitary} shows the contributions of
each of the terms in $\nu_\mu\rightarrow\nu_e$ appearance
for the typical LMA number discussed before.
Because contribution of these terms are all significant,
each of the components in Eq~(\ref{eqn:nueapp})
can be determined by measuring the oscillation pattern of the
$\nu_\mu\rightarrow \nu_e$ and $\bar\nu_\mu\rightarrow \bar\nu_e$
appearances.
This oscillation pattern provides 4
independent measurements
of the MNS matrix elements and can overconstrain
the unitary triangle:
$U^*_{e1}U_{\mu 1}+U^*_{e2}U_{\mu 2}+U^*_{e3}U_{\mu 3}=0$~\cite{JSatoCP}.

\subsection{Sensitivity of proton decay}
\hspace*{\parindent}
The existence of the neutrino oscillation indicates the extremely
small neutriino masses, which is 12-13 orders of magnitude
smaller than the top quark mass. A natural way to explain this
hierarchy is Grand Unified theories (with see-saw mechanism),
which is also indicated by the running gauge coupling constants.
The only known way to directly observe the grand
unification phenomenon is a nucleon decay measurement.
The main gauge-boson-mediated decay is \peppo{} and the
predicted lifetime could be as short as \(\sim 10^{35}\)
years \cite{marciano}.
For supersymmetric grand unified theories,
\pnukp{} decay tends to be the main decay mode
and its predicted life time is somewhere between
$10^{32}-10^{35}$ years, although this decay mode
is highly model dependent. For both of these modes,
it is highly desirable to reach the sensitivity of $\sim$10$^{35}$~years and beyond.
Current lower limits on partial lifetimes of these two modes
from 79 kton-year of  \Super-K{} data \cite{proton.decay.ichep} are
\[\tau_p/B_{\mpeppo} > 5.0 \times 10^{33} \mbox{ years (90\% confidence level)}\] 
\[\tau_p/B_{\mpnukp} > 1.6 \times 10^{33} \mbox{ years (90\% confidence level)}\]
where \(\tau_p\) is the proton lifetime and \(B\) is the branching ratio 
of the decay mode.

\begin{figure}[!tb]
\renewcommand{\baselinestretch}{1}
\centerline{
\epsfig{file=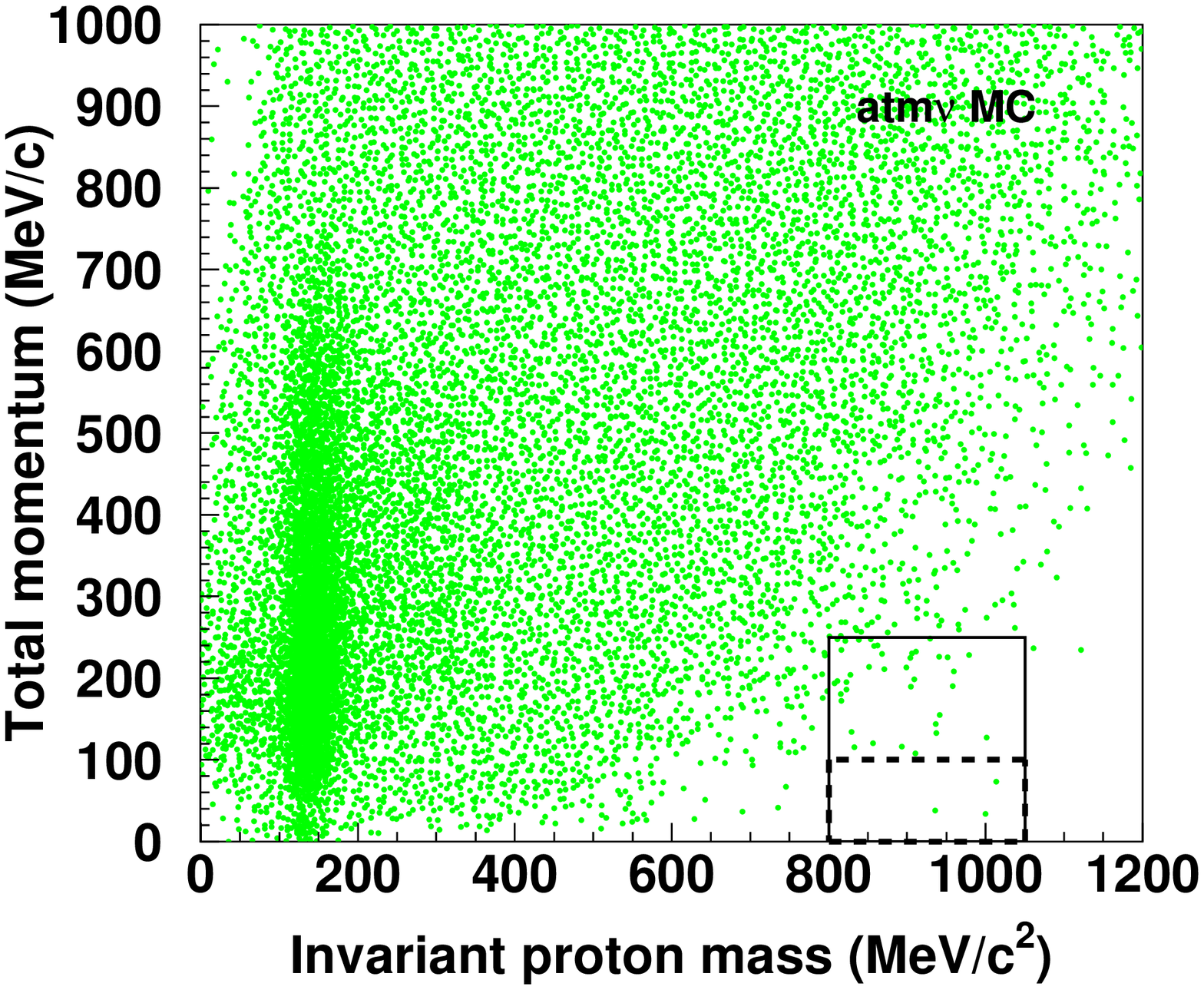,width=0.5\textwidth}
\epsfig{file=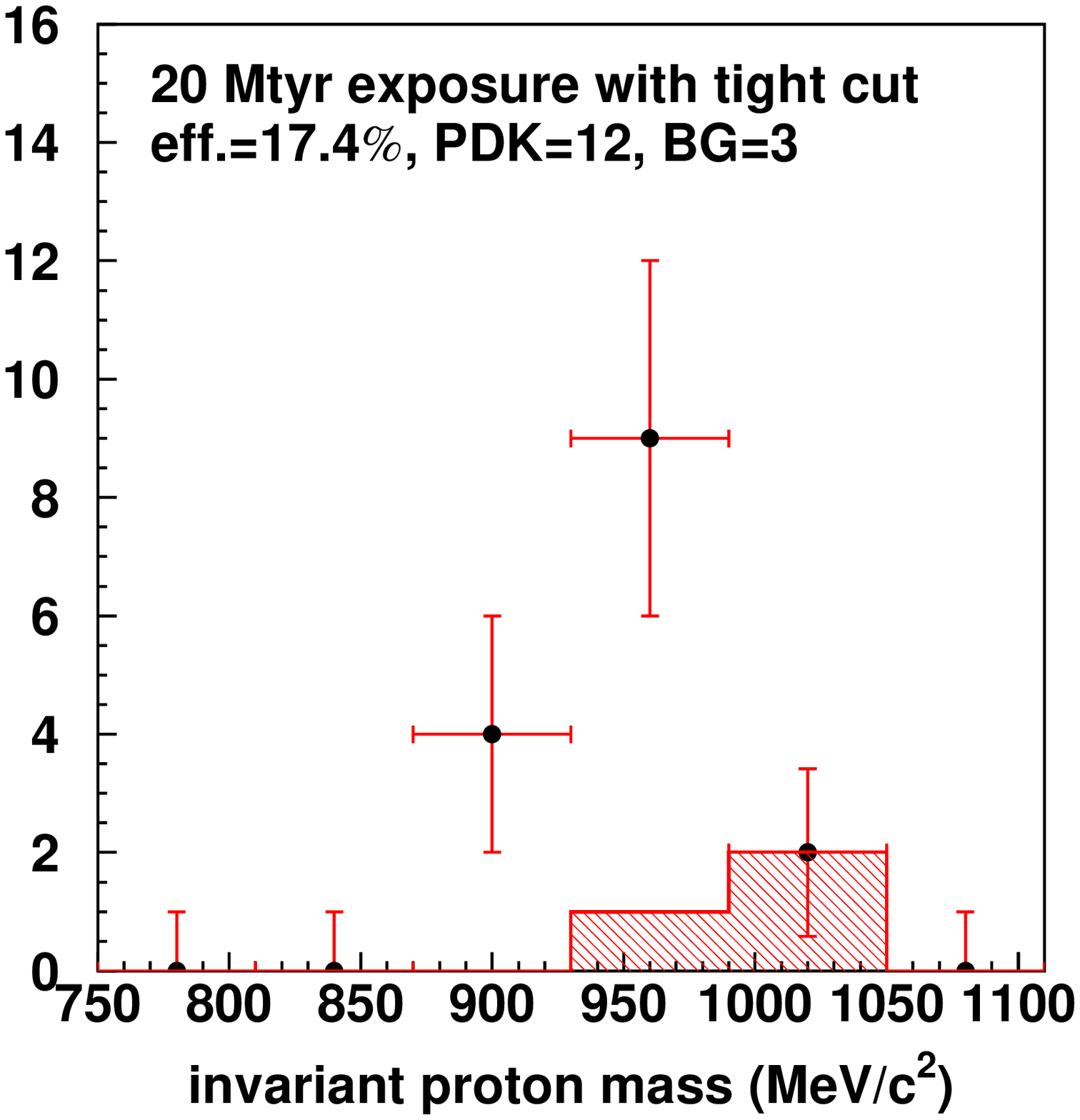,width=0.4\textwidth}
}
\vspace{-3mm}
\caption{\protect\footnotesize
Left:
Observed invariant proton mass and total momentum distributions for
simulated atmospheric neutrino backgrounds of 20~Mt\(\cdot\)year
exposure.  Solid box shows traditional selection criterion used in
\Super-K{} \cite{proton.decay.SK:Phys.Rev.Lett.:1998}.
Dashed box shows new tighter cut for reducing background.
Right: Observed invariant proton mass distributions for
20~Mt\(\cdot\)year exposure.  Partial proton lifetime for
\peppo{} is assumed to be $1 \times 10^{35}$ years.
}
\label{fig:epi0-bg}
\end{figure}

In the following sensitivity study \cite{simulation}, 
the neutrino interaction simulation and detector 
simulation used in \Super-K{} are used.
Figure \ref{fig:epi0-bg} shows simulated atmospheric neutrino backgrounds 
for 20~Mt\(\cdot\)year exposure.  The solid box shows the signal
region used in \Super-K{} \cite{proton.decay.SK:Phys.Rev.Lett.:1998}.  
An estimated number of background events is 45 for
20~Mt\(\cdot\)year and it appears that background limits the statistics
in the future \peppo{} search and a tighter cut is desired.

\begin{figure}[!tb]
\renewcommand{\baselinestretch}{1}
\centerline{
\epsfig{file=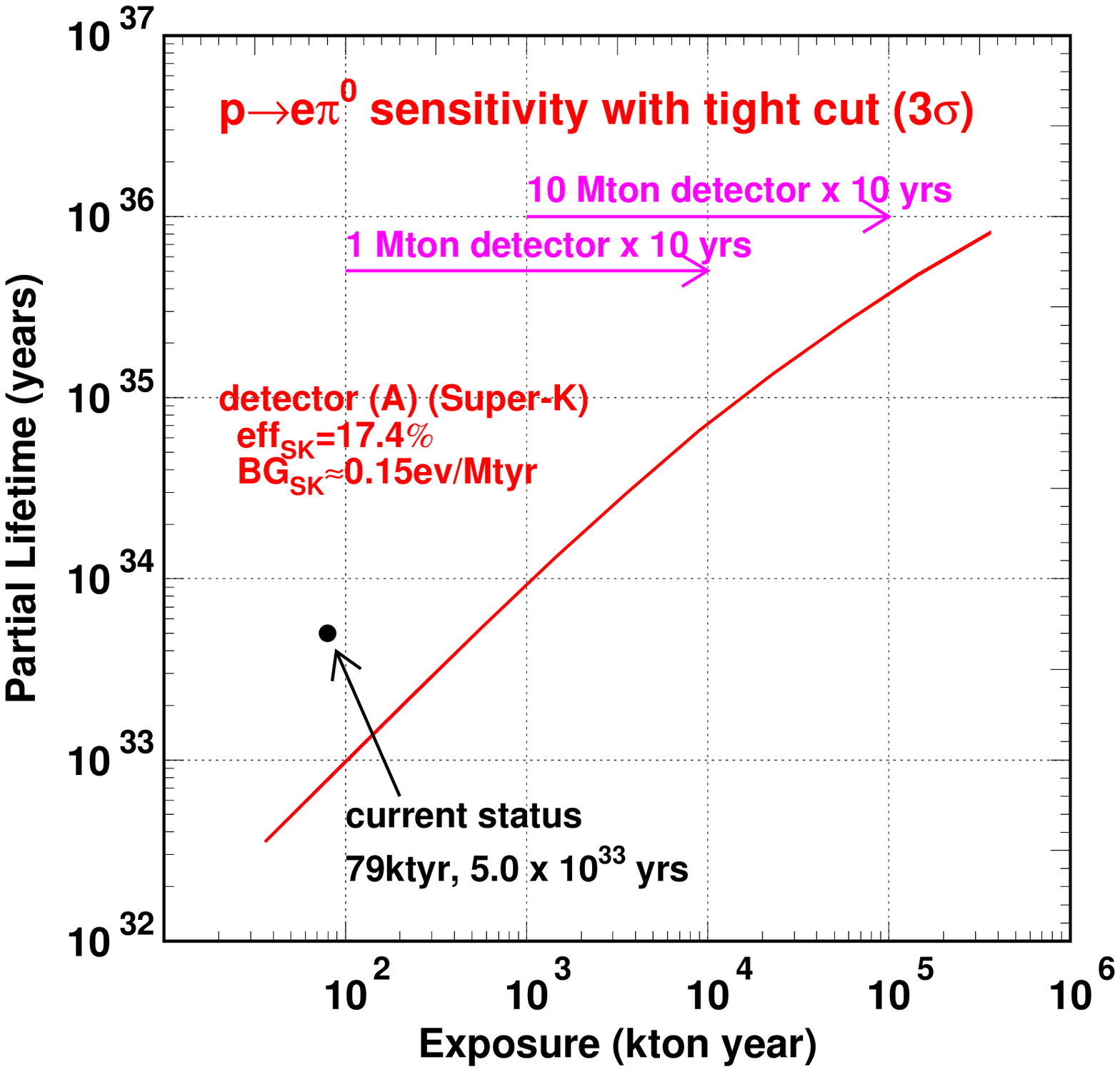,width=0.43\textwidth}
\epsfig{file=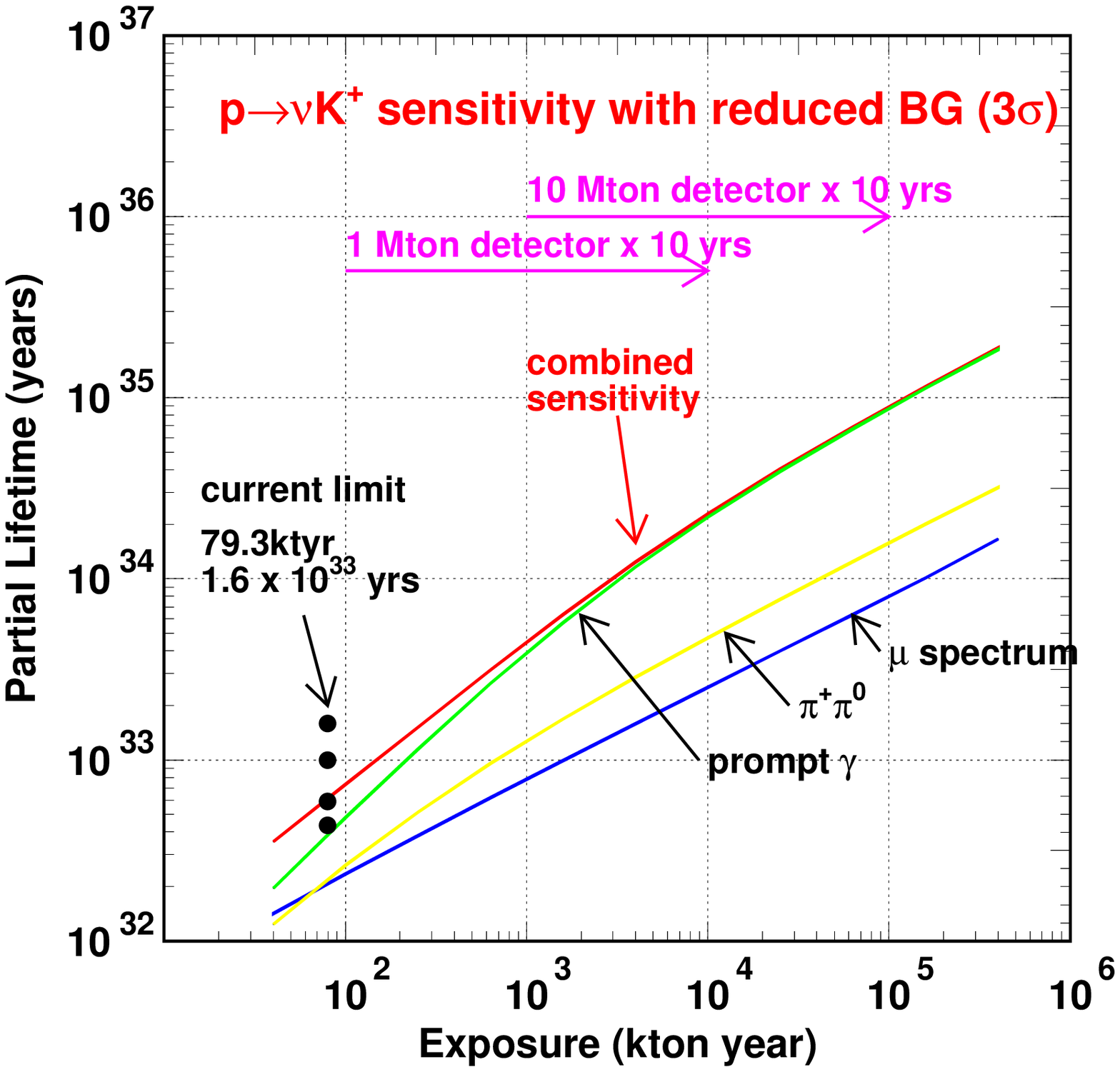,width=0.43\textwidth}
}
\vspace{-4mm}
\caption{\protect\footnotesize
Expected sensitivity for the partial lifetime of protons
as a function of detector exposure.
In the left figure, the sensitivity for \peppo{} is calculated
at 99.73\%(3$\sigma$) confidence level.  The tight momentum cut (see text)
is used in the calculation.  The right figure shows the expected
sensitivity for \pnukp{} mode.  Background is assumed to be reduced 
for muon and prompt gamma tagging method (see text).  The upper line
shows the combined sensitivity for this decay mode.}
\label{fig:sesitivity}
\end{figure}

In the water molecule, 2 out of 10 protons are free protons.
These free protons have no Fermi motion and thus give sharp
proton mass peak in $e^+\pi^0$ invariant mass distribution
(x-axis) and nearly perfect momentum balance (y-axis).
The detection efficiency is also higher because of no pion
interaction loss in the Oxygen nucleus.
The dashed box in the figure represents a tighter
selection criterion in momentum balance
to select only the free proton decay.
The background level is reduced by a factor of 15,
or 3 background events/20~Mt\(\cdot\)year),
whereas the 39\% 
of the signal detection efficiency is maintained.
The overall signal detection efficiency is 17.4\%.
Figure \ref{fig:epi0-bg}(right) shows the expected invariant
mass distribution for 20~Mt\(\cdot\)year exposure data, assuming
a partial lifetime for the proton of \(1 \times 10^{35}\) years and
the tight cut described above.
A sharp peak at the proton mass is seen, which would provide a
redundant positive evidence of proton decay.
Figure \ref{fig:sesitivity}(left) shows 99.73\%(3$\sigma$) discovery
sensitivity. With 20~Mt\(\cdot\)year exposure, we will reach
a sensitivity beyond \(10^{35}\)years.

For the \pnukp{} search, \Super-K{} developed
a nearly background free method of detecting
$\mu^+$ from a $K\rightarrow \mu\nu$ decay accompanied by
a prompt $\gamma$ from the residual oxygen nucleus
\cite{proton.decay.SK:Phys.Rev.Lett.:1999}.
The backgrounds for this prompt $\gamma$ tagging is assumed
to come from kaon production by atmospheric neutrinos.
Figure \ref{fig:sesitivity}(right) shows 99.73\%(3$\sigma$) discovery
sensitivity for the \pnukp{} search.
With 20~Mt\(\cdot\)year exposure, we will reach
a sensitivity of \(3\times 10^{34}\)years,
closing (or opening?) the windows for many
of the supersymmetric grand unified theories.

\section{Summary}
\hspace*{\parindent}
The next generation long baseline neutrino experiment from
JHF to Super Kamiokande with a baseline length of 295~km
is proposed to
explore the physics beyond the Standard Model.
The JHF PS is designed to be capable of
delivering $3.3\times 10^{14}$ protons every 3.42 seconds (0.77~MW),
which is upgradeable to more than five times greater
intensity (4~MW) later.
The high intensity neutrino beam is produced with a
conventional method.
Low energy narrow band beam whose peak energy is tuned to 
the oscillation maximum of $\sim$0.8~GeV is used to maximize
the sensitivity on the neutrino oscillation.
Super-Kamiokande has very a good energy resolution 
and excellent particle identification at this low energy. 
The neutrino energy is reconstructed
through quasi-elastic interaction, which is a dominant
process for the neutrino energy of below 1~GeV.
The experiment will be divided into two phases.
The goals for the first phase are precision measurements of
oscillation parameters in $\nu_\mu$ disappearance, search
for $\nu_e$ appearance, a confirmation of $\nu_{\mu}\rightarrow\nu_{\tau}$ 
oscillation by detecting neutral current events, and 
search for sterile neutrinos.
The expected precision and reach are estimated by using
1) the neutrino flux and spectrum obtained by the simulation code 
of neutrino beam production used by the K2K experiment,
2) Super-Kamiokande's full detector simulator which has been
tested in detail over the last $\gtrsim$5 years for the atmospheric
neutrino observation.
The expected precision is $\delta(\sin^2\!2\theta_{23})\sim 
0.01$, $\delta(\Delta m^2_{23})\lesssim 1\times
10^{-4}$~eV$^2$ in $\nu_{\mu}$ disappearance
and $\nu_e$ appearance can be explored down
to $\sin^2\!2\theta_{\mu e} \sim 3 \times 10^{-3}$.
In the second phase of the experiment with an upgraded 4 MW PS
and 1~Mt Hyper-Kamiokande, $\nu_e$ appearance can be searched
for in the region $\sin^22\theta_{13}<5\times 10^{-4}$ and  CP 
violation can be observed if $|\delta|\gtrsim 10-20^\circ$ in 
the case of large mixing angle solution of the solar neutrinos.
Sensitivity in the proton
decay search is significantly improved 
upto $10^{35}$ ($3\times\!10^{34}$) years in life time for the
$p\rightarrow e^+\pi^0$ ($p\rightarrow \bar{\nu}K^+$) mode.
The first phase experiment is planned to start in 2007.

\end{document}